\documentclass[12pt, a4paper]{article}

\pdfoutput=1

\usepackage{jheppub}
\usepackage{amsmath}	
\usepackage{amssymb}	
\usepackage{bm}		
\usepackage{color} 		
\usepackage{graphicx} 	
\usepackage{times} 		
\usepackage[all]{hypcap}  
\usepackage[sort&compress]{natbib}
\usepackage{hyperref} 	

\hypersetup{
    colorlinks=true,		
    linkcolor=blue,		
    citecolor=green,		
    filecolor=magenta,	
    urlcolor=cyan		
}

\newcommand{\abs}[1]{\left| #1 \right|}

\newcommand{\SU}[1]{\mathrm{SU}\bigl(#1\bigr)}
\newcommand{\U}[1]{\mathrm{U}\bigl(#1\bigr)}
\newcommand{\SO}[1]{\mathrm{SO}\bigl(#1\bigr)}
\newcommand{\eref}[1]{\ensuremath{\mathrm{Eq.}\;(\ref{#1})}}
\newcommand{\fref}[1]{Fig.~\ref{#1}}
\newcommand{\sref}[1]{Sec.~\ref{#1}}
 
\newcommand{\TeV}{\ \mathrm{TeV}}
\newcommand{\GeV}{\ \mathrm{GeV}}
\newcommand{\MeV}{\ \mathrm{MeV}}

\newcommand{\ord}[2][ ]{\mathcal{O}\left( #2 \right)^{#1}}
\newcommand{\ex}[1]{\ \mathrm{exp} \left[ #1 \right]}

\newcommand{\per}{\, .}
\newcommand{\com}{\, ,}
\newcommand{\hc}{\mathrm{h.c.}}

\newcommand{\const}{\mathrm{const.}}

\newcommand{\nn}{\nonumber \\}

\newcommand{\BRinv}{{\rm BR}_{\rm inv}}
\newcommand{\atCL}{\ \text{at} \ 95\% \, {\rm CL}}

\title{The 125 GeV Higgs and Electroweak Phase Transition Model Classes}

\author[a]{Daniel J. H. Chung,}
\author[a, b]{Andrew J. Long,}
\author[c,d]{and Lian-Tao Wang}

\affiliation[a]{Department of Physics, University of Wisconsin, Madison, WI 53706}
\affiliation[b]{Department of Physics and School of Earth and Space Exploration, Arizona State University, Tempe, AZ 85827-1404}
\affiliation[c]{Enrico Fermi Institute and Department of Physics, University of Chicago, Chicago, IL 60637}
\affiliation[d]{Kavli Institute for Cosmological Physics, University of Chicago, Chicago, IL 60637}

\emailAdd{danielchung@wisc.edu}
\emailAdd{andrewjlong@asu.edu}
\emailAdd{liantaow@uchicago.edu}

\abstract{ Recently, the ATLAS and CMS detectors have discovered a
  bosonic particle which, to a reasonable degree of statistical
  uncertainty, fits the profile of the Standard Model Higgs.  One
  obvious implication is that models which predict a significant
  departure from Standard Model phenomenology, such as large 
  exotic (e.g., invisible) Higgs 
  decay or mixing with a hidden sector scalar, are already ruled out.
  This observation threatens the viability of electroweak baryogenesis, 
  which favors, for example, a lighter Higgs and a Higgs
  coupled to or mixed with light scalars.  To assess the broad impact
  of these constraints, we propose a scheme for classifying models of
  the electroweak phase transition and impose constraints on a
  class-by-class basis.  We find that models, such as the MSSM, which
  rely on thermal loop effects are severely constrained by the
  measurement of a 125 GeV Higgs.  Models which rely on tree-level
  effects from a light singlet are also restricted by invisible decay
  and mixing constraints.  Moreover, we find that the parametric
  region favored by electroweak baryogenesis often coincides with an enhanced symmetry
  point with a distinctive phenomenological character.  In particular,
  enhancements arising through an approximate continuous symmetry are
  phenomenologically disfavored, in contrast with enhancements from
  discrete symmetries.  We also comment on the excess of diphoton
  events observed by ATLAS and CMS.  We note that although Higgs
  portal models can accommodate both enhanced diphoton decay and
  a strongly first order electroweak phase transition, the former favors a negative
  Higgs portal coupling whereas the latter favors a positive one, and
  therefore these two constraints are at tension with one another.  }


\begin{document}
\maketitle

\setlength{\parindent}{20pt}
\setlength{\parskip}{2.5ex}

\section{Introduction}
A number of baryogenesis mechanisms are capable of explaining the observed baryon asymmetry of the universe, but many of these operate at a high scale -- inaccessible to direct laboratory tests -- where they evade independent confirmation.  
The primary motivation for studying electroweak baryogenesis (EWBG) \cite{Kuzmin:1985mm} is that the baryon asymmetry is generated by electroweak scale physics, which is tested by experiments aimed at understanding the nature of electroweak symmetry breaking.  
These include Higgs searches at LEP, the Tevatron, and the LHC colliders.  
Thus, models of the electroweak sector may be constrained from two sides: by the requirement that electroweak baryogenesis successfully generates the baryon asymmetry and by the requirement that models remain consistent with Higgs search constraints.  
Indeed, the ATLAS and CMS collaborations recently announced the discovery of a particle in the mass range $125-126 \GeV$ which matches the profile of the Higgs boson  \cite{CMS:2012gu, ATLAS:2012gk}.  
Even at this early stage, without a precise knowledge of the alleged Higgs' couplings to Standard Model (SM) fields, we have gained a partial picture of the origin of the electroweak symmetry breaking (EWSB).  
In this paper, we would like to understand what is the main implications of a $125 \GeV$ SM-like Higgs for electroweak baryogenesis.

Studies of the viability of electroweak baryogenesis and the impact of collider constraints are usually performed on a model-by-model basis.
However, many individual models can accommodate a partial picture of the electroweak symmetry breaking sector.  
Thus, as the LHC begins to expose the Higgs sector, revealing only glimpses of the full picture, one would like to understand what {\it classes of models} may be consistent with or at tension with the data.  To this end, we propose a scheme for classifying models of the electroweak sector based upon the nature of the electroweak phase transition (EWPT) and study the implications of the recent Higgs discovery at the LHC on a class-by-class basis.  
We find that the LHC's detection of a $125 \GeV$ Higgs in conjunction with constraints on exotic decay and hidden sector mixing provide strong constraints on certain EWPT model classes.\footnote{
However, it may be possible to weaken the tension between the Higgs mass measurement and the baryon asymmetry washout condition in nonstandard cosmologies \cite{Servant:2001jh,Barenboim:2012nh}.  
}

\begin{figure}[t]
\begin{center}
\includegraphics[width=0.95\textwidth]{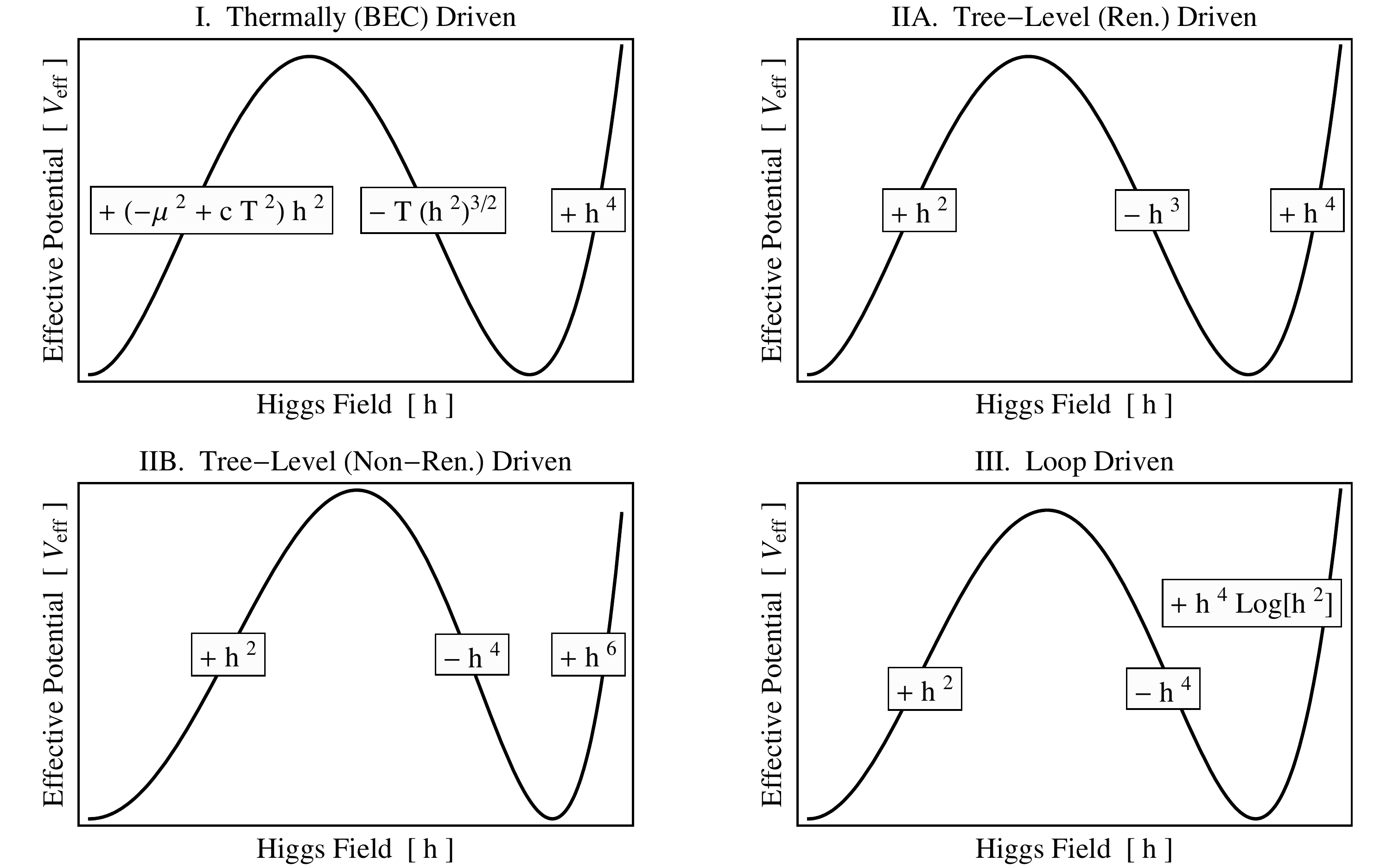}
\caption{\label{fig:Veff_plots} 
The four methods of obtaining a strongly first order phase transition by inducing a barrier in the thermal effective potential, which are discussed in this paper.  
The framed expressions indicate which term is responsible for the rise or fall of $V_{\rm eff}$.  
}
\end{center}
\end{figure}

We identify the phase transition model classes in the following way.
The success of EWBG relies upon the electroweak phase transition being of the first order\footnote{The first order phase transition is a necessary but not sufficient condition.  Additionally, nonequilibrium transport of CP-violating sources is required, and bounds on electric dipole moments lead to strong constraints \cite{Cirigliano:2009yt, Konstandin:2005cd}, which are complimentary to the Higgs constraints discussed herein.  } \cite{Kuzmin:1985mm}.
In the context of the phase transition calculation, this translates into the requirement that the thermal effective potential, $V_{\rm eff}(h,T)$, possesses a pair of minima {\it separated by a barrier} for some range of temperatures \cite{Quiros:1999jp}.  
Thus, we can classify models of the electroweak (EW) sector based on what physics is responsible for providing the requisite barrier in $V_{\rm eff}(h,T)$.  
When calculated perturbatively, $V_{\rm eff}(h , T)$ is given by a sum of tree-level, quantum (loop), and thermal contributions.  
Thus, three model classes can be identified\footnote{
We do not claim that this classification scheme is exhaustive.  Models which cannot be classified in this way include those models which rely on nonperturbative effects (e.g., \cite{Konstandin:2010cd, Konstandin:2011dr}), models for which the relevant physics cannot be qualitatively captured by the high temperature expansion (e.g., \cite{Carena:2004ha}), and models with a nonequilibrium entropy production that cannot be studied in the effective potential formalism.  However, this classification does cover most perturbative models in the literature known to us.  
} (see also \fref{fig:Veff_plots}):
\begin{description}
	\item[\quad {\bf I.  Thermally (BEC) Driven.}]  {\small A
          barrier arises due to thermal loop effects associated with
          bosonic zero modes.  The effective potential acquires a term
          which ideally has the form $-T (h^2)^{3/2}$ where $h$ is the
          Higgs condensate.  Because the nonanalyticity can be traced
          to the lowest energy mode of the Bose-Einstein distribution,
          this can also be intuitively called the Bose-Einstein
          condensation (BEC) driven scenario.  The non-analytic term competes with
          the $h^2$ and $h^4$ terms in the scalar potential to
          generate a barrier.  }
	\item[\quad {\bf II.  Tree-Level Driven.}]  {\small A barrier
          arises due to a competition between terms in the effective
          potential which are already present at tree-level.  This
          model class can be further subdivided.  }
	\item[\qquad \, {\bf IIA.  Renormalizable Operators.}]
          {\small The barrier arises from the competition between
            renormalizable operators.  Since an effective $h^3$
            operator and gauge invariance are required for this class
            of models, these models rely upon a scalar field (or
            fields) in addition to a single Higgs doublet acquiring a
            nonzero expectation value during the EWPT.  }
	\item[\qquad \, {\bf IIB.  Non-Renormalizable Operators.}]
          {\small If nonrenormalizable operators involving the Higgs
            field (such as $h^6$) are added to the scalar potential, a
            barrier can arise as a result of their competition with
            the renormalizable terms.  }
	\item[\quad {\bf III.  Loop Driven.}]  {\small Some $\hbar$
          loop corrections may generate qualitatively important
          nonpolynomial field dependence and aid in generating the barrier.
          For example, Ref.~\cite{Espinosa:2007qk} utilizes the
          quartic correction of the form $h^4 \ln h^2$, which can
          compete with the naively unstable $-h^4$ term to generate a
          barrier.  }
\end{description}

In addition to a barrier in $V_{\rm eff}$, successful EWBG requires the EW sphaleron process to be out of equilibrium in the broken phase to ensure that the baryon asymmetry is not washed out.  
This condition is expressed as a bound on the EWPT order parameter
\cite{Shaposhnikov:1991cu}
\begin{align}\label{eq:SFOcondit}
	\frac{v(T_c)}{T_c} \gtrsim 1.3
\end{align}
where $\left< H \right>_T = \left( 0 \, , \, v(T) / \sqrt{2} \right)^{\rm T}$ is the expectation value of the Higgs at temperature $T$, and $T_c$ is the temperature at which the phase transition takes place (i.e., the symmetric and broken phases have degenerate free energy densities).  
We say that phase transitions which satisfy \eref{eq:SFOcondit} are ``strongly'' first order phase transitions (SFOPT).\footnote{
The exact numerical value of the right hand side of \eref{eq:SFOcondit} is mildly model-dependent, but in all known cases it is a number close to unity.
}

Thus, we will study the EWPT in the context of each model class by first parametrizing the approximate thermal effective potential $V_{\rm eff}$ appropriate for each model class and then investigating what parametric limit will yield $v(T_c) / T_c \gg 1$.  
We can then ask what underlying physics would give rise to such an ``optimal limit,'' what does the associated phenomenology look like, and what is the impact of collider constraints, {\em assuming} that the last Higgs-sector-related phase transition is the electroweak symmetry breaking phase transition (i.e.~there were no phase transitions that jumped from one electroweak symmetry breaking vacuum to another electroweak symmetry breaking vacuum). 
One of the conclusions of our study is  that the optimal limits frequently correspond to enhanced symmetry
points in the theory space.  
This makes the optimal limits straightforward to identify and associates them with a distinctive phenomenology which is constrained by recent LHC data.

For example, one of the EWBG parametric regions most cleanly ruled out by the 125 GeV Higgs is the enhanced {\em continuous} symmetry point parametric region (as opposed to the enhanced {\em discrete} symmetry point), which is a subset of Class IIA (Tree-Level Renormalizable Operator Driven) models.  
As emphasized in Ref.~\cite{Barger:2011vm}, strong first order phase transitions can generically be found near parametric regions surrounding an enhanced symmetry point where the symmetry transformations mix Higgs and another field degree of freedom.  
One subset of enhanced symmetries is based on continuous symmetries (or the parametric limit in which the discrete symmetry enlarges into a continuous symmetry).  
One way to understand how the Higgs data rules out this subset is to note that the Nambu-Goldstone bosons associated with the spontaneously broken continuous symmetries have couplings to Higgs determined by the kinetic part of the action, and this coupling-induced decay rate is unsuppressed when the Higgs mass is of the order of $v=246 \GeV$.  
Hence, the Higgs decay to the Nambu-Goldstone bosons exceeds the experimental limits on exotic decays of the Higgs.

The tension that we present in most of the categorization points to the enhanced discrete symmetry point \cite{Barger:2011vm} being the parametric space marker having intuitively the largest set of model building possibilities for electroweak baryogenesis.

In addition to constraints coming from the SM-likeness of the Higgs, it is also interesting to consider the ``anomalies'' which may point to beyond-the-Standard-Model (BSM) physics.  
One of the most promising anomalies observed at the LHC is an excess of events in the loop-induced diphoton decay channel of the Higgs.  
If the excess can be attributed to the presence of a BSM scalar field running in the loop, then we utilize our classification to argue that there is a general tension with electroweak baryogenesis if this scalar field is also responsible for driving a SFOPT.  

The order of presentation is as follows.  
We begin with a review of the collider data relevant for analysis.  
In Sec.~\ref{sec:EWPTModelClasses}, we present our classification and a general discussion of how the current data affects the models in the classification.  
We also give explicit model examples that fit into the proposed classification.  
In Sec.~\ref{sec:diphoton}, we discuss the impact of the diphoton excess anomaly on each of the classes.  We then close the paper with a conclusion.

\section{Collider Data and Interpretation}\label{sec:DataandInterpretation}

Since models of the EW sector with strongly first order EW phase
transitions tend to rely on a large coupling between the Higgs and
light scalar fields, it is important to review the relevant
constraints here.  The Tevatron signal and ATLAS/CMS discovery confirm
the existence of a 
bosonic particle 
with an approximate mass of $125 \GeV$
\cite{CMS:2012gu, ATLAS:2012gk}.  
The available statistics
suggest that the decays of this boson are consistent with the SM
predictions in the channel $b \bar{b}$
\cite{Tevatron:2012cn,Abazov:2012xx,Tevatron:2012sx} as well as $ZZ \to 4 \ell$ and
$WW \to \ell \nu \ell \nu$ 
\cite{CMS:2012gu, ATLAS:2012gk}.
In the diphoton decay channel, both ATLAS and CMS observe
an excess of events above the SM prediction
\cite{CMS:2012gu, ATLAS:2012gk}.

{\bf Spectrum:} It is well known that in models such as the SM and the
Minimal Supersymmetric Standard Model (MSSM), even the LEP Higgs mass
bound imposes strong constraints on the viability of EWBG.  Of course,
these constraints have already ruled out EWBG in the SM
\cite{Kajantie:1996mn}.  
The measurement of a Higgs mass of $125 \GeV$ further severely restricts the allowed MSSM parameter space \cite{Curtin:2012aa, Cohen:2012zz}, although EWBG in MSSM is still viable with more judicious choices of parameters \cite{Carena:2012np}.

{\bf Exotic (e.g., Invisible) Decay:} 
The discovery of a SM-like Higgs at the LHC is at tension with a large branching fraction in exotic channels.  
For instance, if the Higgs had a large branching fraction to invisibles, $\BRinv = {\rm BR}(h \to {\rm inv})$, this would suppress the branching fraction in all visible channels, and it would have been more difficult to find the Higgs at the LHC\footnote{
Assuming that the new physics does not enhance the Higgs production cross section, i.e., we assume $\sigma(pp \to h) = \sigma^{\rm SM}(pp \to h)$. However, even in the MSSM where new physics both allows invisible decay and enhances the Higgs production cross section, one finds that invisible decay is at tension with the data \cite{Desai:2012qy}.  } \cite{Barger:2008jx}.  
A number of groups have investigated this possibility by assuming that the production cross section is the same as for a $125 \GeV$ SM Higgs, but allowing for $\BRinv$ to vary in fitting the data.
They obtain upper bounds on the branching fraction to invisibles in the range $\BRinv < 0.30-0.75 \atCL$ \cite{Giardino:2012ww,Englert:2012ha,Espinosa:2012vu,Carmi:2012in,Espinosa:2012im,Giardino:2012dp,Montull:2012ik}.
Although this may not seem overly restrictive, we will see that in the phase transition model classes which allow invisible decay, this is naturally the dominant decay channel.
Furthermore, the LHC expects to resolve the issue of invisible decay with increased data.  
It is estimated that with $20 \, {\rm fb}^{-1}$ integrated luminosity, the LHC should detect or exclude invisible decay for $\BRinv > 0.4 \atCL$ \cite{Bai:2011wz}, and at $30 \, {\rm fb}^{-1}$, ATLAS should detect or exclude invisible decay for $\BRinv > 0.24$ at $5 \sigma$ \cite{Barger:2008jx}.
Partially invisible final states resulting from exotic cascade decays are more difficult to constrain, but branching fraction bounds on the order of $10 \%$ may be obtained at the LHC with $1000 \, {\rm fb}^{-1}$ \cite{Englert:2012at}.  

{\bf Mixing with Hidden Sector:}  
Just as with the case of invisible decay, the ATLAS/CMS data
strongly constrains the scenario in which the Higgs is allowed to mix
with a hidden sector scalar field or fields, which are singlets under
the SM gauge group.  For the sake of discussion, we will suppose that
only one singlet scalar field is mixing with the SM Higgs.  The impact
of this mixing on the phenomenology depends on the relative mass
scales, of the Higgs-like scalar at $m_H \approx 125 \GeV$ and the
singlet-like scalar with mass $m_{\rm hid}$.  Let $\theta$ be the
angle between the Higgs-like mass eigenstate and the Higgs gauge
eigenstate.  The relevant constraints are: \\
\parbox{5.7in}{
\begin{enumerate}
	\item {\bf Light Higgs search at LEP.}  {\small The existence
          of a light singlet-like resonance (i.e., $m_{\rm hid} \ll
          m_H = 125 \GeV$) is constrained by Higgs searches at LEP.
          In order for the singlet-like particle to have evaded
          detection, its coupling to the SM must be suppressed.  This
          places an upper bound on $\theta$, which becomes more
          stringent as $m_{\rm hid}$ is decreased below the LEP Higgs
          search bound of $114.4 \GeV$.  For instance, for $m_{\rm
            hid} = 20 \GeV$ one needs $\cos^2 \theta > 0.99 \atCL$
\cite{Barate:2003sz}.  }
	\item {\bf Heavy Higgs search at LHC.}  {\small Similarly, if
          the singlet-like resonance is heavier (i.e., $m_{\rm hid}
          \gg m_H = 125 \GeV$), there is an upper bound on $\theta$
          coming from the requirement that the heavy singlet-like
          scalar evades detection at the LHC.  Again, this is a
          function of the singlet-like scalar's mass.  For instance,
          if $m_{\rm hid} = 200 \GeV$ one needs
          $\cos^2 \theta > 0.60 \atCL$ to avoid detection
          \cite{Azatov:2012bz,Bai:2011aa}.  }
	\item {\bf LHC Higgs Detection.}  {\small 
Assuming that the Higgs-like resonance is lighter (i.e., $m_H = 125 \GeV \ll m_{\rm hid}$), then the consequence of mixing is a universal suppression of all Higgs production processes by a factor of $\cos^2 \theta$.  
Large mixing would have made discovery more difficult.  
Thus, the LHC's signal at $125 \GeV$ places an upper bound on $\theta$ which may be expressed as $\cos^2 \theta > 0.77$ at $90\%$ CL \cite{Carmi:2012yp, Giardino:2012dp}. (See also \cite{Giardino:2012ww,Azatov:2012bz,Azatov:2012rd,Espinosa:2012ir,Bertolini:2012gu}).
        }
\end{enumerate}
}\\ Taken together, these constraints imply that the large-mixing
scenario (e.g., $\cos^2 \theta = 0.5$) is strongly disfavored.

\section{Electroweak Phase Transition Model Classes}\label{sec:EWPTModelClasses}

In this section, we will enumerate the phase transition model classes,
identify the parametric limits which are optimal for SFOPT by
maximizing the washout criterion \eref{eq:SFOcondit}, and discuss
phenomenological constraints that arise in those limits.  To connect
to phenomenological constraints, we will make a simplifying assumption
that the electroweak symmetry breaking is the {\it last} Higgs-sector-related 
phase transition (i.e., there is no transition from one electroweak
nonsymmetric vacuum to another electroweak nonsymmetric
vacuum).\footnote{With a sufficiently large number of broken vacua
  jumps in between, the electroweak symmetry breaking vacuum
  properties measured at colliders today can be decoupled from the
  vacuum properties associated with the first electroweak symmetry
  breaking phase transition.}  As we discuss further below, the
optimal limits for SFOPT often correspond to enhanced symmetry points
of the theory at which the symmetry group is extended to include an
additional continuous or discrete symmetry.  For the sake of brevity,
we will not dwell on the details of the phase transition calculation.
We refer the interested reader to the review \cite{Quiros:1999jp}.

\subsection{Class I: Thermally (BEC) Driven}\label{subsec:I}
 
In models such as the SM and the MSSM, the barrier in the thermal
effective potential arises from thermal loop effects, which emerge in
the following way.  The Higgs condensate $\left< H \right> = \left( 0
\, , h / \sqrt{2} \right)^{T}$ modifies the dispersion relation of
particles in the plasma causing them to acquire an effective
temperature- and field-dependent mass $m_{\rm eff}^2(h,T) =
\tilde{m}^2(h) + \Pi(T)$.  Here, $\Pi$ is a temperature-dependent
self-energy correction (known as ``daisy resummation,'' see e.g.,
\cite{Carrington:1991hz}) and $\tilde{m}^2(h)$ can be obtained by
replacing the zero temperature {\sc vev} $v$ with $h$ in the standard
expression for the field's mass (see e.g., \cite{Quiros:1999jp}).
Bosonic fields induce a contribution to the thermal effective
potential of the form $V_{\rm eff} \ni (-T/12 \pi) \bigl( m_{\rm
  eff}^2(h,T) \bigr)^{3/2}$ in the high-temperature limit.  The
nonanalyticity of this term at $m_{\rm eff}^2=0$ can be traced to the
nonanalyticity of the Bose-Einstein distribution function at zero
energy.  Hence, this thermal ``BEC term''-driven SFOPT defines our
``Class I'' model class.

To achieve a barrier in $V_{\rm eff}$ near the phase transition 
temperature $T_c$, we want to have $(m_{\rm eff}^2(h, T_c))^{3/2} \sim h^3$ 
such that there may be a competition between this term and the $h^2$ 
and $h^4$ terms of the Higgs potential.  
Supposing that $\tilde{m}^2(h)$ can be written as
$\tilde{m}^2(h) = \alpha h^2 + \beta$, the effective mass will have
the desired scaling if we tune $\beta - \Pi(T_c) \ll \alpha v(T_c)^2$.
A general phenomenological consequence of this tuning is that the
scalar bosons today will be light, since their mass squared is
$\tilde{m}^2(v) \approx \alpha v^2 - \Pi(T_c)$.  Note that increasing
the interaction of the $h$ field to make $\alpha$ large naturally
drives up $\Pi(T_c)$ which in turn drives $\tilde{m}^2(v)$ lighter.
The need for this tuning using $\beta$ is well-established in the MSSM
\cite{Carena:1996wj, Delepine:1996vn}, where light right-handed stops provide this
$(m_{\rm eff}^2(h, T_c))^{3/2}$ term.  
Phenomenologically, the light stops tend to enhance Higgs production by gluon fusion and reduce Higgs diphoton decay.  
Because of this, the LHC has already placed strong constraints on EWBG in the MSSM \cite{Curtin:2012aa, Cohen:2012zz}, and it has begun to push the model into a corner that will be probed by the high luminosity LHC \cite{Carena:2012np}.

Near the temperature of the phase transition, the effective potential
may be approximated as
\begin{align}\label{eq:TDriven_Veff}
	V_{\rm eff}(h, T) \approx \frac{1}{2} \left( - \mu^2 + c \, T^2 \right) h^2 - \frac{e \, T}{12 \pi} (h^2)^{3/2} + \frac{\lambda}{4} h^4
\end{align}
in the high temperature expansion.  Note that a factor of $1/12\pi$
has been included in the parametrization to reflect the natural
thermal loop suppression of this coefficient.  A potential of this
form is illustrated in \fref{fig:Veff_plots}.  The parameters $\mu^2 =
m_H^2 / 2$ and $\lambda = m_H^2 / (2 v^2)$ are related to the Higgs
mass $m_H$ and {\sc vev} $v$.\footnote{More generally, $h$ need not be the
  Higgs and $m_H$ need not be the $125 \GeV$ Higgs mass.  }.  The
dimensionless parameters $c$ and $e$ quantify the coupling between the
Higgs condensate and the relativistic particles in the plasma.  In
particular, $c$ depends on couplings between $h$ and light ($m < T$)
bosons and fermions, whereas $e$ only depends upon couplings between
$h$ and light bosons.  Schematically, 
\begin{align}
	e \sim \sum_{\text{light bosonic fields}} (\text{degrees of freedom}) \times (\text{coupling to Higgs})^{3/2} \per
\end{align}
The contribution from heavy fields ($m > T$)
are Boltzmann-suppressed, and the $O(T^4 \ex{-m/T})$ terms are
dropped.  Some examples of models that fall into this class are shown
in Table \ref{tab:TDriven_Examples}\footnote{In models such as the MSSM and ``Colored Scalar'' model, the light scalars that provide the BEC term are colored (e.g., stops in the MSSM).  
Two-loop QCD corrections to the effective potential may strengthen the phase transition by up to an $O(1)$ factor \cite{Espinosa:1996qw, Cohen:2011ap}.  
We do not incorporate two-loop corrections into our analysis as we expect the qualitative parametric behavior to be dominantly controlled by the leading-order terms.  
}.

\begin{table}[t] \tiny
\hspace{-2in}
\vspace{-0in}
\begin{center}
\begin{tabular}{|p{2cm}|p{3.0cm}|c|p{4cm}|}
\hline
Model 
	& $-\Delta \mathcal{L}$
	& $\boldsymbol{c}$ 
	& $\boldsymbol{e}$ \\ \hline \hline
SM \cite{Anderson:1991zb} 
	& 
	& $c_{\rm SM} = \frac{6 m_t^2 + 6 m_W^2 + 3 m_Z^2 + \frac{3}{2} m_H^2}{12 v^2}$ 
	& $e_{\rm SM} = \frac{6 m_W^3 + 3 m_Z^3}{v^3}$   \\ \hline
MSSM \cite{Carena:1996wj}
	& 
	& $c_{\rm SM} + \frac{6m_t^2}{12v^2} \left( 1 - \frac{\tilde{A}_t^2}{m_Q^2} \right)$
	& $e_{\rm SM} + \frac{6m_t^3}{v^3} \left( 1 - \frac{\tilde{A}_t^2}{m_Q^2} \right)^{3/2}$  \\ \hline
Colored Scalar \cite{Cohen:2012zz}
	& $M_X^2 \abs{X}^2 + \frac{K}{6} \abs{X}^4 + Q \abs{H}^2 \abs{X}^2$
	& $c_{\rm SM} + \frac{6}{24} \frac{Q}{2}$
	& $e_{\rm SM} + 6 \left( \frac{Q}{2} \right)^{3/2}$  \\ \hline
Singlet Scalar \cite{Anderson:1991zb, Espinosa:1993bs} 
	& $M^2 \abs{S}^2 + \lambda_S \abs{S}^4 + 2 \zeta^2 \abs{H}^2 \abs{S}^2$
	& $c_{\rm SM} + \frac{g_S}{24} \zeta^2$ 
	& $e_{\rm SM} + g_S \zeta^3$  \\ \hline
Singlet \quad Majoron \cite{Cline:2009sn}
	& $\mu_s^2 \abs{S}^2 + \lambda_s \abs{S}^4 + \lambda_{hs} \abs{H}^2 \abs{S}^2 + \frac{1}{2} y_i S \nu_i \nu_i + \hc$
	& $c_{\rm SM} + \frac{2}{24} \frac{\lambda_{hs}}{2}$ 
	& $e_{\rm SM} + 2 \left( \frac{\lambda_{hs}}{2} \right)^{3/2}$  \\ \hline
Two Higgs Doublets \cite{Chowdhury:2011ga} 
	& $\mu_D^2 D^{\dagger}D + \lambda_D (D^{\dagger}D)^2 + \lambda_3 H^{\dagger} H D^{\dagger}D + \lambda_4 \abs{H^{\dagger}D}^2 + (\lambda_5/2) [(H^{\dagger}D)^2 + \hc ]$ 
	& $c_{\rm SM} +  \frac{2 \lambda_3 + \lambda_4}{12}$
	& $e_{\rm SM} + 2 \left( \frac{\lambda_3}{2} \right)^{3/2} + \left( \frac{ \lambda_3 + \lambda_4 - \lambda_5}{2} \right)^{3/2} + \left( \frac{\lambda_3 + \lambda_4 + \lambda_5}{2} \right)^{3/2}$ \\ \hline
\end{tabular}
\end{center}
\caption{
\label{tab:TDriven_Examples}
Examples of models in the Thermally (BEC) Driven class. 
The expressions for $e$ are calculated in the limit that the field-independent contributions to $m_{\rm eff}^2(h,T)$ are negligible (e.g., the thermal mass tuning has been performed).  
Here, the symbol $\tilde{A}_t$ is $\tilde{A}_t=A_t -\mu/\tan \beta$ and $g_s$ is the number of real scalar singlet degrees of freedom coupling to the Higgs.  
}
\end{table}%

A standard calculation (see, e.g., Ref.~\cite{Quiros:1999jp}) yields the
EW order parameter
\begin{align}\label{eq:TDriven_voverT}
	\frac{v(T_c)}{T_c} \approx \frac{e}{6 \pi \lambda} \, .
\end{align}
There are two ``optimal'' limits in which we can obtain $v(T_c) / T_c
\gg 1$.
\begin{description}
	\item[${\bm e} \, {\bm \gg} \, {\bm \lambda}$] \qquad To reach
          the limit of large $e$, the Higgs must have a large coupling
          with many light bosonic degrees of freedom.  Indeed, the
          presence of $6\pi$ in the denominator of
          Eq.~(\ref{eq:TDriven_voverT}) (which comes from the thermal
          loop expansion) makes satisfying Eq.~(\ref{eq:SFOcondit})
          very challenging if $\lambda\sim O(1)$.  There are various
          phenomenological constraints on this limit.  First, since
          $e$ is a sum of dimensionless coupling constants (see, e.g.,
          Table \ref{tab:TDriven_Examples}), it is bounded from above
          by the perturbative unitarity constraint.  Second, heavy
          bosonic fields will become Boltzmann-suppressed and cannot
          contribute to $e$.  However, the same interactions which
          allow light bosonic fields to contribute to $e$ also provide
          a mass to those fields after EWSB.  Thus, increasing the
          coupling constants that enter $e$, will eventually cause the
          bosons to become heavy and their contributions to $e$ will become
          Boltzmann-suppressed.\footnote{Note that the examples
            we have chosen in Table \ref{tab:TDriven_Examples} do not
            include $\mathbb{Z}_2$-breaking 
            cubic couplings since those would naturally have strong
            phase transition possibilities driven by tree-level
            terms.} 
(One can however increase $e$
          up to the perturbativity bound by increasing the number of
          degrees of freedom that contribute to the thermal loop
          instead of increasing the coupling constant.  However, in
          that case, one may need to arrange fermion loops to cancel
          radiative corrections to $\lambda$.)
Finally, as $e$ is increased, interactions between the Higgs and other
bosonic fields are made stronger.  Thus, there may be loop-suppressed --
but nevertheless significant -- modifications to Higgs production and /
or decay.  For example, if the bosons carry color, then they can
significantly enhance Higgs production by gluon fusion
\cite{Cohen:2012zz,Buckley:2012em}.  We will revisit this constraint
in the context of Higgs diphoton decay in \sref{sec:diphoton}.
	\item[${\bm \lambda} \, {\bm \to} \, {\bm 0}$] \qquad In the
          context of the SM, this limit is obviously forbidden in
          light of the relationship $\lambda = m_H^2 / 2 v^2$ and the
          fact that $m_H$ is now a measured quantity.  However, in an
          effort to keep our model classification scheme as general as
          possible, we will consider the scenario in which the field
          $h$ that appears in \eref{eq:TDriven_Veff} is not the SM
          Higgs condensate.  Instead, it may represent a
          parametrization of some non-trivial trajectory through an
          extended scalar field space connecting the EW-preserving
          vacuum $h=0$ with the EW-broken vacuum $h=v$.  Then, the
          limit $\lambda \to 0$ implies the spectrum contains a light
          scalar.  
          If the scalar carries SM quantum numbers, then
          direct search constraints are severe unless the scalar can
          be hidden in a large SM background.  If the Higgs decay channel
          is open, this limit may be at tension with constraints on Higgs exotic 
          decay.  If the scalar is a SM singlet, then constraints on hidden sector mixing
          may also apply.  
Moreover, vacuum stability considerations limit the range of the EFT (see, e.g., \cite{EliasMiro:2011aa} and references therein).  
\end{description}

To illustrate how these limits and constraints arise in a concrete model, we extend the SM by a color triplet scalar field $X$ (see \cite{Cohen:2012zz}):  
\begin{align}
	\mathcal{L} = \mathcal{L}_{\rm SM} +  \left( \partial_{\mu} X \right)^{\ast} \left( \partial^{\mu} X \right) - \left[ M_X^2 X^{\ast} X + \frac{K}{6} \left( X^{\ast} X \right)^2 + Q H^{\dagger} H X^{\ast} X \right] \per
\end{align}
Note that the quantum numbers of $X$ only allow it to couple to the EW sector 
via the so-called ``Higgs portal'' operator $H^2 X^2$ with coefficient $Q$.  
The effective mass of the scalar $X$ is given by $m_X^2(h,T) = M_X^2 + (Q/2) h^2+\Pi_X(T)$ where $\Pi_X(T) = (K+Q) T^2/24$.  
Thus, the BEC term is given by 
\begin{align}\label{eq:TDriven_Veff_example}
	\Delta V_{\rm eff}(h, T) = - 6 \frac{T}{12 \pi} \left( M_X^2 + \frac{Q}{2} h^2 + \Pi_X(T) \right)^{3/2} \, ,
\end{align}
where the factor $6$ is the number of internal degrees of freedom for the complex, colored $X$ field.  
As discussed above, we must tune $M_X^2 \approx -\Pi_X(T_c)$.  Thus, \eref{eq:TDriven_Veff_example} takes the form of \eref{eq:TDriven_Veff} with $e = e_{\rm SM} + 6 (Q/2)^{3/2}$.  

\begin{figure}[t]
\begin{center}
\includegraphics[width=0.70\textwidth]{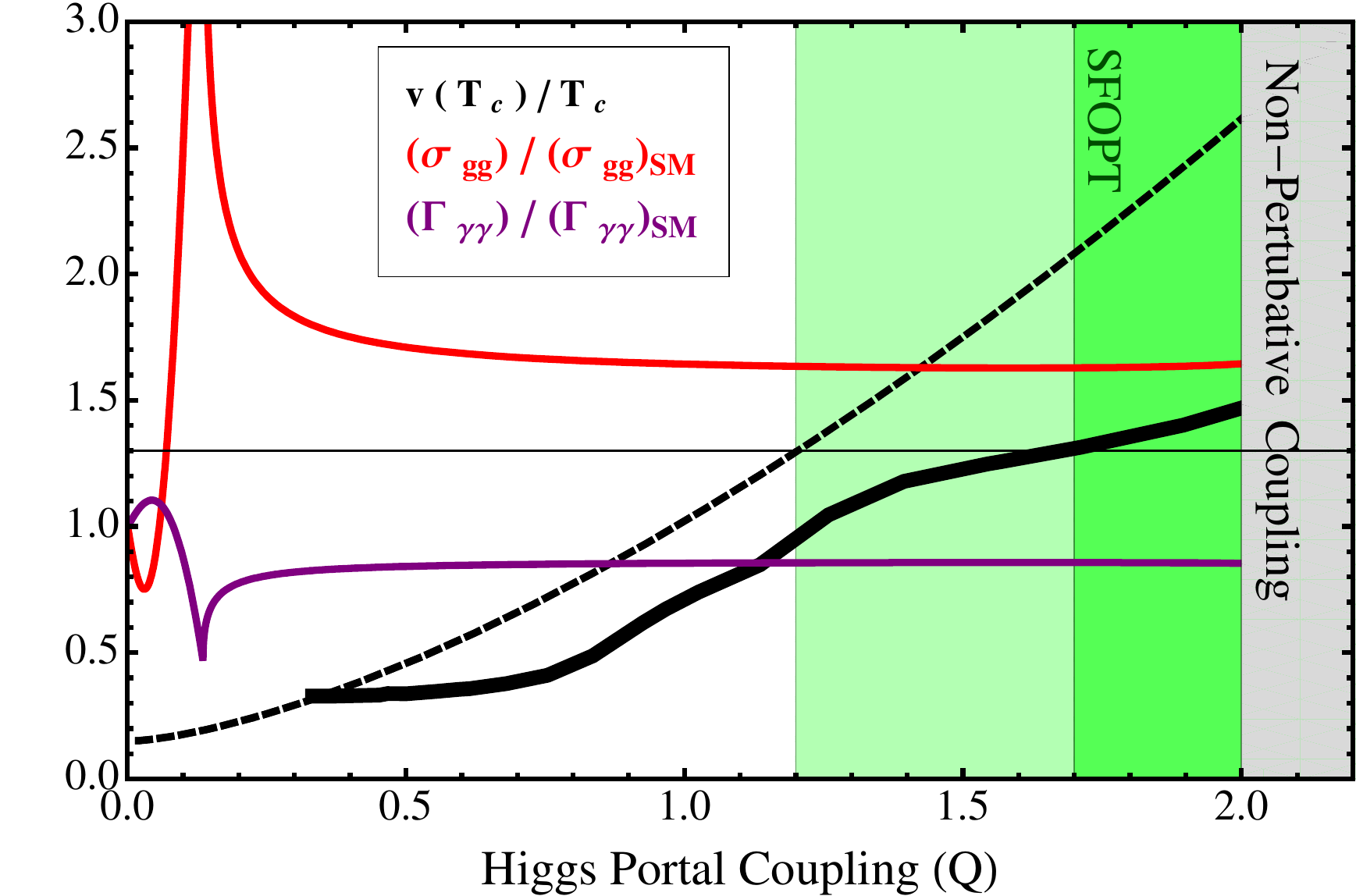} 
\caption{\label{fig:TDriven_orderparam}  A plot of the EW order
  parameter $v(T_c)/T_c$ calculated analytically (black, dashed) and numerically 
  (black, solid), as discussed in the text, as well as modifications to
  Higgs production by gluon fusion (red) and Higgs decay to two
  photons (purple).  
  The numeric calculation of $v(T_c) / T_c$ falls short of the analytic
  estimate due to the Boltzmann suppression effect discussed in the
  text.  The analytic expression suggests that SFOPT are obtained for
  $Q \gtrsim 1.2$, but numerical calculation reveals that SFOPT are
  only found for $Q \gtrsim 1.7$.  Hence, there is a narrow window
  $1.7 \lesssim Q \lesssim 2.0$ where the perturbative calculation is
  valid and the EWPT is strongly first order.  
In this region, when $X$ is an electrically charged color triplet, the phenomenology consists of an enhanced rate of $gg \to H$ and a reduced $H \to \gamma \gamma$ rate.  
}
\end{center}
\end{figure}

We would like to understand what constraints arise as we go to the
SFOPT limit $e \gg \lambda$.  This limit is reached by taking $Q \gg
\lambda^{2/3}$.  First, we verify that the phase transition is
strongly first order by calculating $v(T_c) / T_c$ as a function of
$Q$.  We fix $m_H = 125 \GeV$, $K = 0.1$, and require $M_X^2 = - \Pi_X(T_c)$ 
where 
\begin{align}
	T_c = \sqrt{\frac{\lambda v^2}{c}} \left[ 1 - \frac{\lambda}{2c}  \left( \frac{e}{6 \pi \lambda} \right)^2 \right]^{-1/2} \per
\end{align}
The numerical calculation is
performed in the standard way (see, e.g., \cite{Quiros:1999jp}) using
the full one-loop, daisy-improved thermal effective potential.  As shown in
\fref{fig:TDriven_orderparam}, the EW phase transition becomes
strongly first order for sufficiently large values of $Q \gtrsim
1.7$. 
Second, we note that perturbativity up to 100 TeV requires $Q < 2$ at the weak scale \cite{Cohen:2012zz}.  
Third, as we discussed above, Boltzmann suppression of heavy $X$ bosons prevents us from obtaining SFOPT for arbitrarily large $Q$.  
We can estimate an upper bound on $Q$ by requiring the $X$ bosons to be light at the temperature of the phase transition:  $T_c > m_X(v(T_c),T_c) \approx \sqrt{Q/2} v(T_c)$ translates into $Q \lesssim 2 (T_c / v(T_c))^2 \approx 2$.  
The numerical calculation confirms this estimate and explains the discrepancy between the numerical an analytic calculations of $v(T_c) / T_c$.  

As noted in Ref.~\cite{Cohen:2012zz}, the addition of a color triplet $X$ field to the SM is motivated by a desire to use the six real scalar degrees of freedom to enhance the strength of the EWPT.  
Such a scenario has a natural connection with collider physics, via the $X$ field's contribution to the rate of Higgs production by gluon fusion (see also \cite{Kribs:2012kz}).  
Furthermore, if the $X$ field is electrically charged like the stops, then it can also affect the diphoton decay rate of the Higgs.
To illuminate this point, we also show in \fref{fig:TDriven_orderparam} the modifications to gluon fusion and diphoton decay, which are calculated following \cite{Batell:2011pz} and using $q_X = 2/3$ for the electric charge of the $X$ field.  
As shown in \fref{fig:TDriven_orderparam}, in the SFOPT window, gluon fusion is enhanced and $H \to \gamma \gamma$ decay is suppressed by an $O(1)$ factor with respect to the SM rates.  
This phenomenological signature is marginally disfavored by the recent ATLAS/CMS data, but additional statistics will be required to justify a strong statement.  

In summary, even if the $X$ fields are SM singlets and thereby able to evade collider restrictions, \fref{fig:TDriven_orderparam} illustrates a generic strong tension from phase transition and theoretical considerations alone.  
The tension is ultimately tied to the difficulty of overcoming the natural $1/6\pi$ suppression appearing in \eref{eq:TDriven_voverT} while remaining in the perturbative regime.  
The most promising way of overcoming this natural suppression is to have a small effective quartic coupling which is model-dependently constrained by collider observations since it typically signals light particle states which have not been observed.

\subsection{Class IIA: Tree-Level (Renormalizable Operators) Driven}\label{subsec:IIA}
We saw in the previous section that the Thermally (BEC) Driven models are strongly constrained, ultimately because of their reliance on the BEC term and its thermal-loop-suppression factor of $1/6\pi$.  
Our next class of models which we call ``Class II'' relies instead on tree-level interactions of the Higgs to provide the barrier for the SFOPT.  
For renormalizable models, these tree-level operators are cubic in the fields (with respect to a particular field origin associated with the EWPT).  
Then, gauge invariance requires that there be at least one scalar in addition to the SM Higgs that acquires an expectation value during the EWPT.  
For nonrenormalizable models, a barrier may be obtained without any odd-powered monomial terms, and therefore we will further subdivide the tree-level model class into two subclasses (``Class IIA'' and ``IIB'').  
We will find perhaps the most clean nontrivial result of this paper that a particular corner of the Class II model class is ruled out due to the current Higgs data.

\begin{table}[t] \small
\begin{center}
\begin{tabular}{|p{4cm}|c|}
\hline
Model 
	& $\Delta \mathcal{L}$ \\ \hline \hline
{\rm xSM} \cite{Ham:2004cf, Profumo:2007wc, Ashoorioon:2009nf, Espinosa:2011ax} 
	& $\frac{1}{2} \left( \partial S \right)^2 - \left[ \frac{b_2}{2} S^2 + \frac{b_3}{3} S^3 + \frac{b_4}{4} S^4 + \frac{a_1}{2} H^{\dagger} H S^2 + \frac{a_2}{2} H^{\dagger} H S^2 \right]$
	\\ \hline
$\mathbb{Z}_2${\rm xSM} \cite{Enqvist:1992va, Barger:2011vm} 
	& $\frac{1}{2} \left( \partial S \right)^2 - \left[ \frac{b_2}{2} S^2 + \frac{b_4}{4} S^4 + \frac{a_2}{2} H^{\dagger} H S^2 \right]$
	\\ \hline
Two Higgs Doublets \cite{Cline:2011mm}
	& $\mu_D^2 \abs{D}^2 + \lambda_D \abs{D}^4 + \lambda_3 \abs{H}^2 \abs{D}^2 + \lambda_4 \abs{H^{\dagger}D}^2 + (\lambda_5/2) [(H^{\dagger}D)^2 + \hc ]$
	\\ \hline \hline
Model 
	& $\Delta W$ \\ \hline \hline
{\rm NMSSM} \cite{Pietroni:1992in, Davies:1996qn, Huber:2000mg} 
	& $\lambda H_1 H_2 N - \frac{\kappa}{3} N^3 + r N$
	\\ \hline
{\rm nMSSM} \cite{Menon:2004wv} 
	& $\lambda H_1 H_2 S + \frac{m_{12}^2}{\lambda} S$
	\\ \hline
{\rm $\mu \nu$MSSM} \cite{Chung:2010cd} 
	& $-\lambda_i H_1 H_2 \nu_i^c + \frac{\kappa_{ijk}}{3} \nu_i^c \nu_j^c \nu_k^c + Y_{\nu}^{ij} H_2 L_i \nu_{j}^c$
	\\ \hline
\end{tabular}
\end{center}
\caption{
\label{tab:CDrivenA_Examples}
Examples of models that fall into Class IIA.  For the non-SUSY models, corrections to the SM Lagrangian are shown, whereas for the SUSY models only the superpotential corrections are given.  
}
\end{table}

First, we consider the class of models (which we call ``Class IIA'') in which the barrier in $V_{\rm eff}$ arises from renormalizable tree-level interactions between the Higgs and new scalar fields.
Thus, the term in $V_{\rm eff}$ that provides the barrier is necessarily cubic, and derived from dimension-three or -four scalar interactions in the Lagrangian.  
This naturally does not suffer from the $1/6\pi$ thermal-loop-factor handicap as in Class I models.  
As we remarked above, at least a single scalar degree of freedom in addition to the SM Higgs must participate in the phase transition.  
We thus parametrize the additional scalar field(s) as $S$.  
The number of degrees of freedom associated with $S$, its quantum numbers, and its interactions will be model-dependent.  
The information that is pertinent to our generic phase transition analysis is that there exists a one-dimensional trajectory through the configuration space which interpolates between the EW-symmetric and EW-broken phases.\footnote{We can parametrize the one-dimensional trajectory with
  a field $\varphi$, as $h = \bar{h}(\varphi, T)$ and $S =
  \bar{S}(\varphi, T)$.  In principle, the functions $\bar{h}$ and
  $\bar{S}$ can be determined by solving for the multifield bounce
  solution.  }  
The effective potential along this trajectory may be approximated as
\begin{align}\label{eq:CDrivenA_Veff}
	V_{\rm eff}(\varphi, T) \approx \frac{1}{2} \left( m^2 + c \, T^2 \right) \varphi^2 - \mathcal{E} \, \varphi^3 + \frac{\lambda}{4} \varphi^4 \, ,
\end{align}
where we have only included the leading high-temperature dependence,
since by definition of this model class, we are assuming that the
temperature-independent tree-level $\varphi^3$ is more important than
the naturally suppressed $T (\varphi^2)^{3/2}/12\pi$ term.  The fact
that this model class does not have to generate a cubic term
dynamically and overcome the natural $1/12\pi$ suppression gives this
class a considerably larger model freedom than Class I.  
Moreover, we neglect the tadpole terms $M^3 \varphi$ and $M T^2 \varphi$, which can be removed by a shift in the origin of the coordinate system.  
Although the unspecified shift obscures the connection between \eref{eq:CDrivenA_Veff} and the underlying theory parameters, we will see that the one-dimensional approximation nevertheless allows us to extract qualitative connections between the phase transition and phenomenology.  
Some examples of models that fall into this class are shown in Table \ref{tab:CDrivenA_Examples}.  

\begin{figure}[t]
\begin{center}
\includegraphics[width=0.45\textwidth]{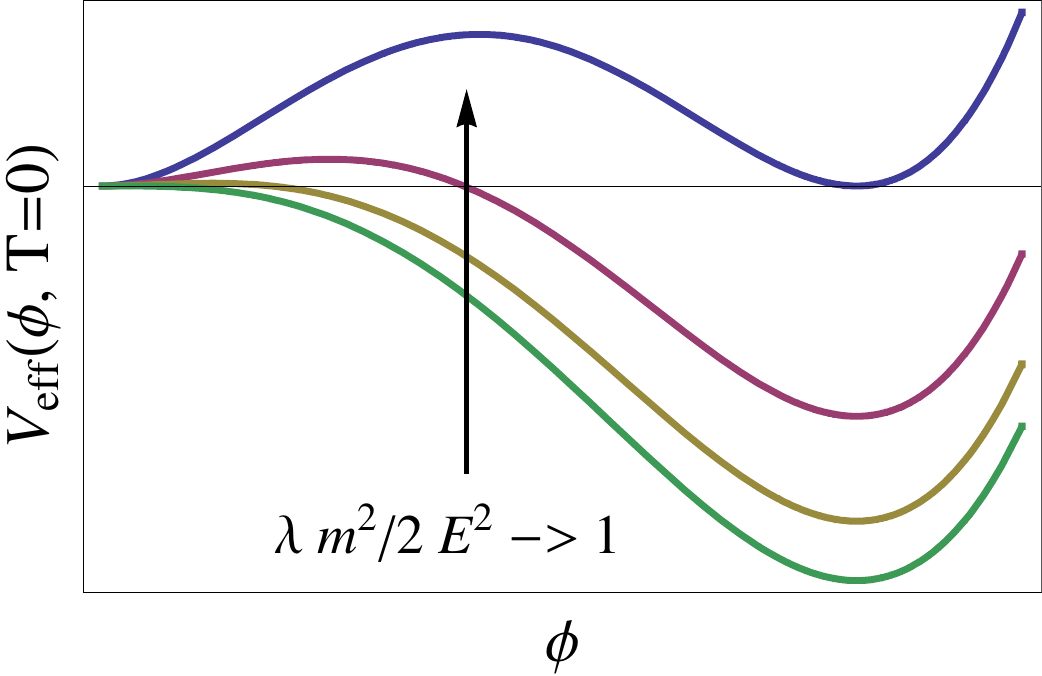} \hfill
\includegraphics[width=0.45\textwidth]{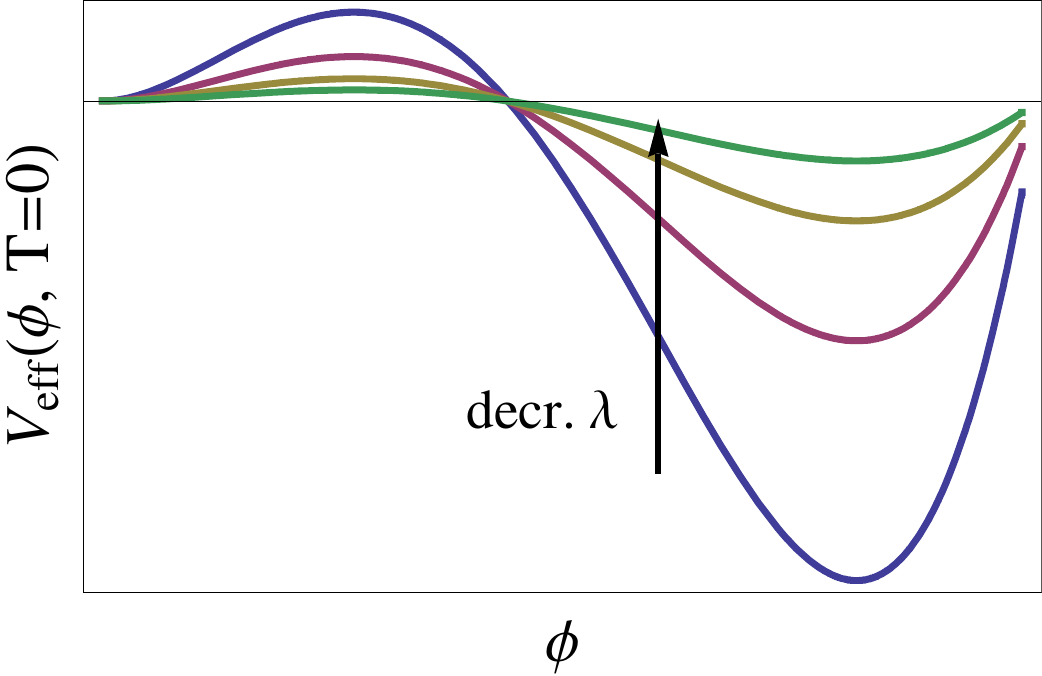} \hfill
\caption{\label{fig:CDriven_limits}
An illustration of the behavior of $V_{\rm eff}$ as the limits $\lambda m^2 / 2 \mathcal{E}^2 \to 1$ (left) and $\lambda \to 0$ (right) are taken.  The former leads to an EdSP whereas the latter leads to an EcSP.  
}
\end{center}
\end{figure}

The phase transition temperature is calculated from \eref{eq:CDrivenA_Veff} to be
\begin{align}\label{eq:CDrivenA_Tc}
	T_c \approx \sqrt{\frac{m^2}{c} } \sqrt{\frac{2 \mathcal{E}^2}{\lambda m^2} - 1} \, ,
\end{align}
and the EW order parameter is found to be
\begin{align}\label{eq:CDrivenA_voverT}
	\frac{v(T_c)}{T_c} \approx \sqrt{ \frac{2c}{\lambda} } \frac{1}{\sqrt{ 1 - \frac{\lambda m^2}{2 \mathcal{E}^2}} } \cos \alpha \, .
\end{align}
Here, we have introduced a projection factor of $\cos \alpha$, since
in general $\varphi$ will not correspond to the Higgs field.
The optimal limits for enhancing $v(T_c) / T_c$ are given by:
\begin{description}
	\item[$\boldsymbol{c \gg \lambda}:$] \qquad Since $c$
          represents a sum of coupling constants controlling
          interactions between the Higgs and light particle in the
          plasma, one might try to take the limit $c \gg \lambda$ by
          increasing the size of these couplings or by increasing the
          number of degrees of freedom in the plasma.  Although this 
          limit is similar to the $e\gg \lambda$ case discussed for
          Class I, they differ significantly in that $e$ only
          receives contributions from bosonic degrees of freedom
          (recall the name BEC Driven), whereas $c$ receives contributions
          from fermions as well.  The Higgs self-coupling $\lambda$ is
          also renormalized by these same couplings that enhance $c$.
          Generally, it is not obvious that the limit used to increase
          $c$ will not also increase $\lambda$ and thereby prevent one
          from reaching the $c \gg \lambda$ limit.  For example, we
          can consider the contributions to $c$ and $\lambda$ that
          arise from the Yukawa interaction with the top quark.  The
          contributions scale with the Yukawa coupling $h_t$ and the
          number of colors $N_c$ like $c \sim N_c h_t^2$ and
          $\lambda \sim -N_c h_t^4$ yielding $c / \lambda \sim -1 /
          h_t^2$.  In this example, increasing the value of the Yukawa
          coupling will tend to decrease the ratio of $c / \lambda$.
          One way to get around this result is to note that
          contributions to $c$ are non-negative whereas contributions
          to $\lambda$ are positive for bosonic fields and negative
          for fermionic fields.  If the underlying model possesses a
          symmetry relating bosonic and fermionic fields (such as
          SUSY) then it may be possible to take $c$ large while
          keeping $\lambda$ small.  If the light fields do not carry
          any SM quantum numbers, and if they are sufficiently light
          ($m < m_H / 2$) then $c \gg \lambda$ is at tension with
          constraints on Higgs invisible decay.
	\item[$\boldsymbol{\lambda m^2 / 2 \mathcal{E}^2 \to 1}:$] \qquad
This is the limit in which $T_c$ vanishes and the EW-symmetric and EW-broken vacua are degenerate.  
As noted in Ref.~\cite{Barger:2011vm}, this degeneracy may arise as the result of a discrete symmetry relating the Higgs field with the other field(s) participating in the phase transition.  
We will refer to this limit as an enhanced (discrete) symmetry point (EdSP), which is illustrated in \fref{fig:CDriven_limits}.  
As one approaches the EdSP, the EW-symmetric vacuum becomes metastable and increasingly degenerate with the EW-broken vacuum.  
Without sufficient degeneracy breaking, tunneling out of the EW-symmetric vacuum may become suppressed to the point that tunneling occurs on a time scale that exceeds the age of the universe.  
That is, as one approaches the EdSP, it may be the case that the EWPT never occurs, even if the EW-broken vacuum is energetically favored.  
	\item[$\boldsymbol{\lambda \to 0}:$] \qquad
We would like to take this limit while fixing $\lambda m^2 / 2 \mathcal{E}^2$ such that \eref{eq:CDrivenA_voverT} just scales like $1/\sqrt{\lambda}$.  
Moreover, if we also want to fix the {\sc vev} of the $\varphi$ field
\begin{align}
	v_{\varphi} = \frac{3\mathcal{E}}{2 \lambda} \left( 1 + \sqrt{1 - \frac{8}{9} \frac{\lambda m^2}{2 \mathcal{E}^2} } \right)
\end{align}
then we see that we must let $\mathcal{E} \propto \lambda$ and $m^2 \propto \lambda$ as $\lambda \to 0$.  
In this limit, the mass of the $\varphi$ field
\begin{align}
	m_{\varphi}^2 = \frac{9 \mathcal{E}^2}{2 \lambda} \left( 1 - \frac{8}{9} \frac{\lambda m^2}{2 \mathcal{E}^2} + \sqrt{1 - \frac{8}{9} \frac{\lambda m^2}{2 \mathcal{E}^2} } \right)
\end{align}
also scales like $\lambda$ and goes to zero.  
Thus, there will be a light scalar field associated with the $\varphi$ field direction.  
The light scalar runs into two phenomenological constraints.  
If $\varphi$ represents a mixture of the Higgs with a hidden sector scalar field, then a light Higgs is excluded by searches at LEP and at tension with the LHC Higgs discovery.  
On the other hand, even if there is no mixing, provided that the light scalar is mostly a SM singlet, then this limit runs into constraints on Higgs invisible decay imposed by the LHC Higgs discovery.  
We will discuss this scenario further in an example below.  

It is important to note that as we take this limit in which $m^2$, $\mathcal{E}$, and $\lambda$ approach zero, the effective potential develops a shift symmetry.  
Thus, we can identify the $\lambda \to 0$ limit with an enhanced symmetry point of the theory at which a continuous symmetry emerges.  
We will refer to this parametric limit as an enhanced (continuous) symmetry point (EcSP), which is illustrated in \fref{fig:CDriven_limits}.  
\end{description}

In order to demonstrate how these limits and constraints may be realized in a concrete model, we consider the $\mathbb{Z}_2$xSM \cite{Barger:2011vm}.  
This model extends the SM by a real scalar field $S$ which is a singlet under the SM gauge group, but which respects a $\mathbb{Z}_2$ discrete symmetry that takes $S \to - S$.  
The most general, renormalizable Lagrangian consistent with the SM gauge group and $\mathbb{Z}_2$ is given by\footnote{Since the one-loop phase transition analysis does not depend upon the quantum  numbers of $S$, the analysis here will also apply to the more general case of a non-singlet $S$ coupled via the ``Higgs portal'' operator $H^{\dagger} H S^{\ast} S$.  Such a scenario is discussed in \sref{sec:diphoton}.  }
\begin{align}\label{eq:L_Z2xSM}
	\mathcal{L}_{\mathbb{Z}_2 {\rm xSM}} = \mathcal{L}_{\rm SM} +  \frac{1}{2} \left( \partial_{\mu} S \right) \left( \partial^{\mu} S \right) - \left[ -\frac{b_2}{2} S^2 + \frac{b_4}{4} S^4 + \frac{a_2}{2} H^{\dagger} H S^2 \right] \, ,
\end{align}
where $\mathcal{L}_{\rm SM}$ is the Lagrangian of the SM.  
We assume that $S$ does not acquire a {\sc vev}.  
Thus the $\mathbb{Z}_2$ is unbroken, thereby ensuring stability of $S$ and preventing mixing with the Higgs.  
Although $S$ does not have a {\sc vev}, we will allow it to obtain a
nonzero expectation value at finite temperature so that it may
participate in the EWPT and render it strongly first order.

With this Lagrangian, we can calculate the effective potential as a function of both the Higgs condensate $\left< H \right> = \left( 0 \, , \, h / \sqrt{2} \right)^{T}$ and singlet condensate $\left< S \right> = s$.   
Working to the same level of approximation as in \eref{eq:CDrivenA_Veff}, we neglect the loop-suppressed contributions and include only the leading thermal contributions to obtain
\begin{align}\label{eq:CDrivenA_Veff_example}
	V_{\rm eff}(h, s, T) = \frac{-\mu^2 + c_h \, T^2}{2} h^2 + \frac{\lambda}{4} h^4 + \frac{-b_2 + c_s \, T^2}{2}  s^2 + \frac{b_4}{4} s^4 + \frac{a_2}{4} h^2 s^2 \, .
\end{align}
The thermal mass terms $c_h \, T^2$ and $c_s \, T^2$ ensure symmetry restoration at sufficiently high temperature.

In light of the general analysis of the preceding subsections, we are motivated to seek out enhanced symmetry points.  
In the following discussion, we will identify the EcSP, justify the claim that SFOPT are found in its vicinity, determine the phenomenology in this limit, and assess the impact of collider constraints.  
We will then repeat the analysis for a neighborhood of the EdSP.

The parameters of the $\mathbb{Z}_2$xSM are the SM gauge ($g_i$) and Yukawa couplings ($y_i$), the Higgs sector parameters ($\mu^2$ and $\lambda$), the singlet sector parameters ($b_2$ and $b_4$), and the ``Higgs portal'' coupling ($a_2$).  
The symmetry group of the $\mathbb{Z}_2$xSM Lagrangian is $G_{\rm SM} \times \mathbb{Z}_2$ where $G_{\rm SM}$ is the gauge group of the SM.  
For a particular choice of parameters, the symmetry group enlarges to incorporate an additional continuous symmetry.  
We find this EcSP by requiring 
\begin{align}\label{eq:EcSP}
	{\rm EcSP:} \quad \left\{ 
	b_2 = \mu^2 \quad  , \quad
	b_4 = \lambda \quad  , \quad
	a_2 = 2 \lambda
	\right\} \quad
	\text{and} \quad  \left\{g_i = 0 \quad , \quad y_i = 0 \right\} ,
\end{align}
where $\lambda = m_H^2 / (2 v^2)$ and $\mu^2 = m_H^2 / 2$ are not constrained by the symmetry, but are restricted by measurements of the Higgs mass and {\sc vev}.  
At the EcSP, the Lagrangian can be written as 
\begin{align}\label{eq:L_Z2xSM_EcSP}
	\mathcal{L}_{\mathbb{Z}_2 {\rm xSM}} \Bigr|_{\rm EcSP} \supset &\left( \partial_{\mu} H \right)^{\dagger} \left( \partial^{\mu} H \right) + \frac{1}{2} \left( \partial_{\mu} S \right) \left( \partial^{\mu} S \right) \nn
	&- \left[ -\mu^2 \left( H^{\dagger} H + S^2 /2 \right) + \lambda \left( H^{\dagger} H + S^2/2  \right)^2 \right]
\end{align}
up to kinetic terms for the other SM fields.  
By virtue of the EW symmetry, this Lagrangian is invariant under an $\SO{4}$ symmetry which acts on the components of $H$.  
However, by virtue of the EcSP, this symmetry is enlarged to an $\SO{5}$ group\footnote{This symmetry relation between the Higgs and singlet fields arises, for example, in nonminimal composite Higgs models \cite{Gripaios:2009pe, Espinosa:2011eu}.  } which rotates among the components of $H$ and $S$.  
The symmetry ensures that $c_s = c_h = c_0$ and the effective potential may be written as 
\begin{align}\label{eq:Veff_EcSP}
	V_{\rm eff}(h,s,T) \Bigr|_{\rm EcSP} = \frac{1}{2} \left(-\mu^2 + c_0 \, T^2 \right) \left( h^2 + s^2 \right) + \frac{\lambda}{4} \left( h^2 + s^2 \right)^2 \, .
\end{align}
Evidently the restriction to vanishing gauge and Yukawa couplings is unphysical, and once these couplings are turned on, radiative corrections to $V_{\rm eff}$ will break the $\SO{5}$ symmetry back down to $\SO{4}$.  
However, the symmetry-breaking terms will carry a loop-suppression factor of $1 / 16 \pi^2$ and can be neglected at this level of approximation.  
On the other hand, contributions to the thermal masses are not loop-suppressed and will generically induce $c_h \neq c_s$.  
Therefore, in the following discussion we will neglect loop-suppressed corrections to $V_{\rm eff}$, but we will treat $c_h$ and $c_s$ as independent parameters.  

We will see that there are SFOPT in a neighborhood of \eref{eq:EcSP},
but first it is interesting to remark that the pattern of symmetry
breaking is controlled by the symmetry that arises at this EcSP.
Provided that $\mu^2 > 0$, the continuous symmetry will be spontaneously broken.
The resulting Nambu-Goldstone boson is associated with a flat
direction in the potential connecting $\abs{H} = v / \sqrt{2}$ with $S
= v$.  Thus, we anticipate that we will find phase transitions that
occur in two steps: first $S$ acquires an expectation value breaking
the $\mathbb{Z}_2$, and second the expectation value of $S$ returns to
zero as $H$ acquires an expectation value breaking the EW symmetry.

\begin{figure}[t]
\begin{center}
\includegraphics[width=0.45\textwidth]{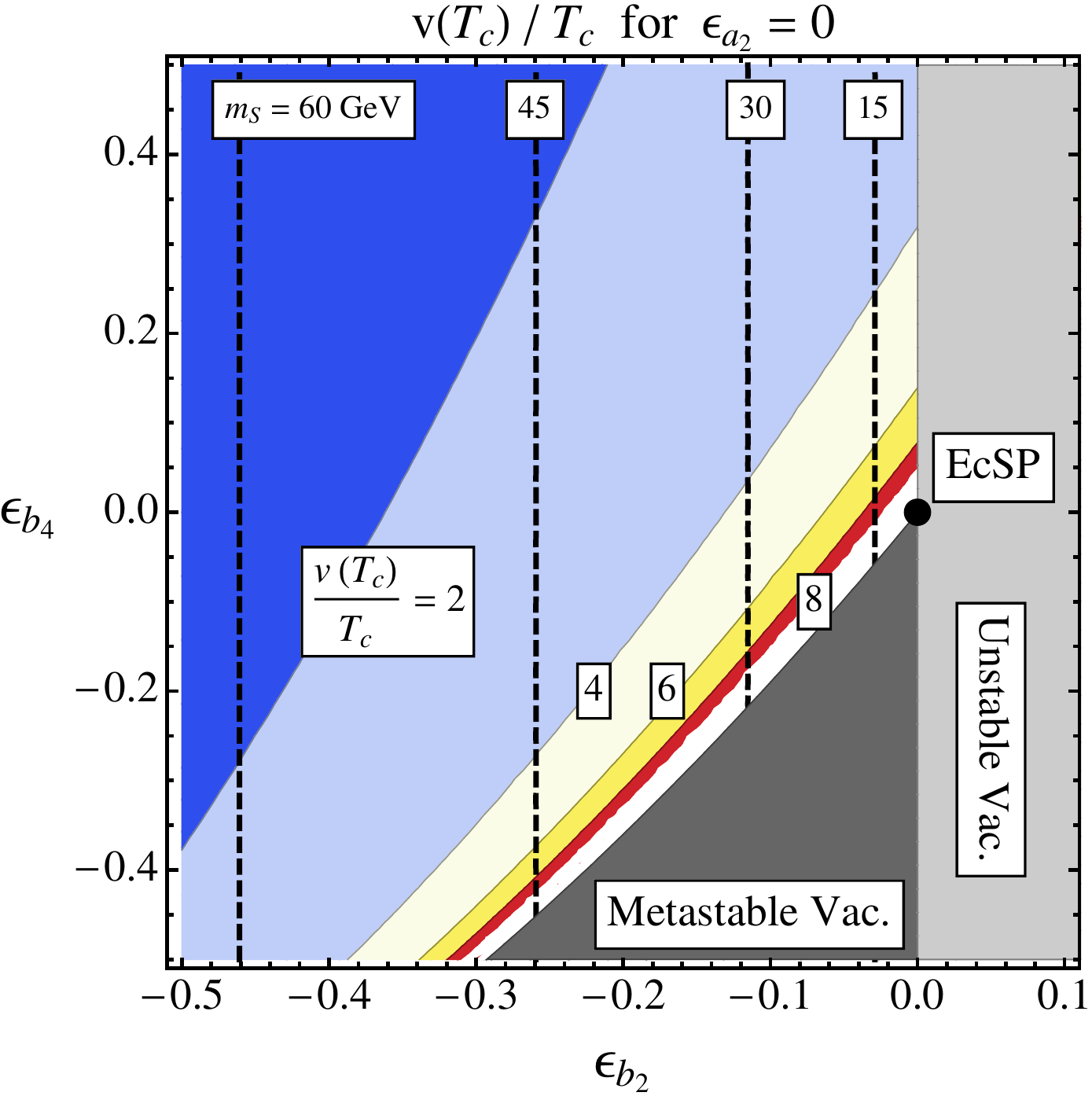}  \hfill 
\includegraphics[width=0.45\textwidth]{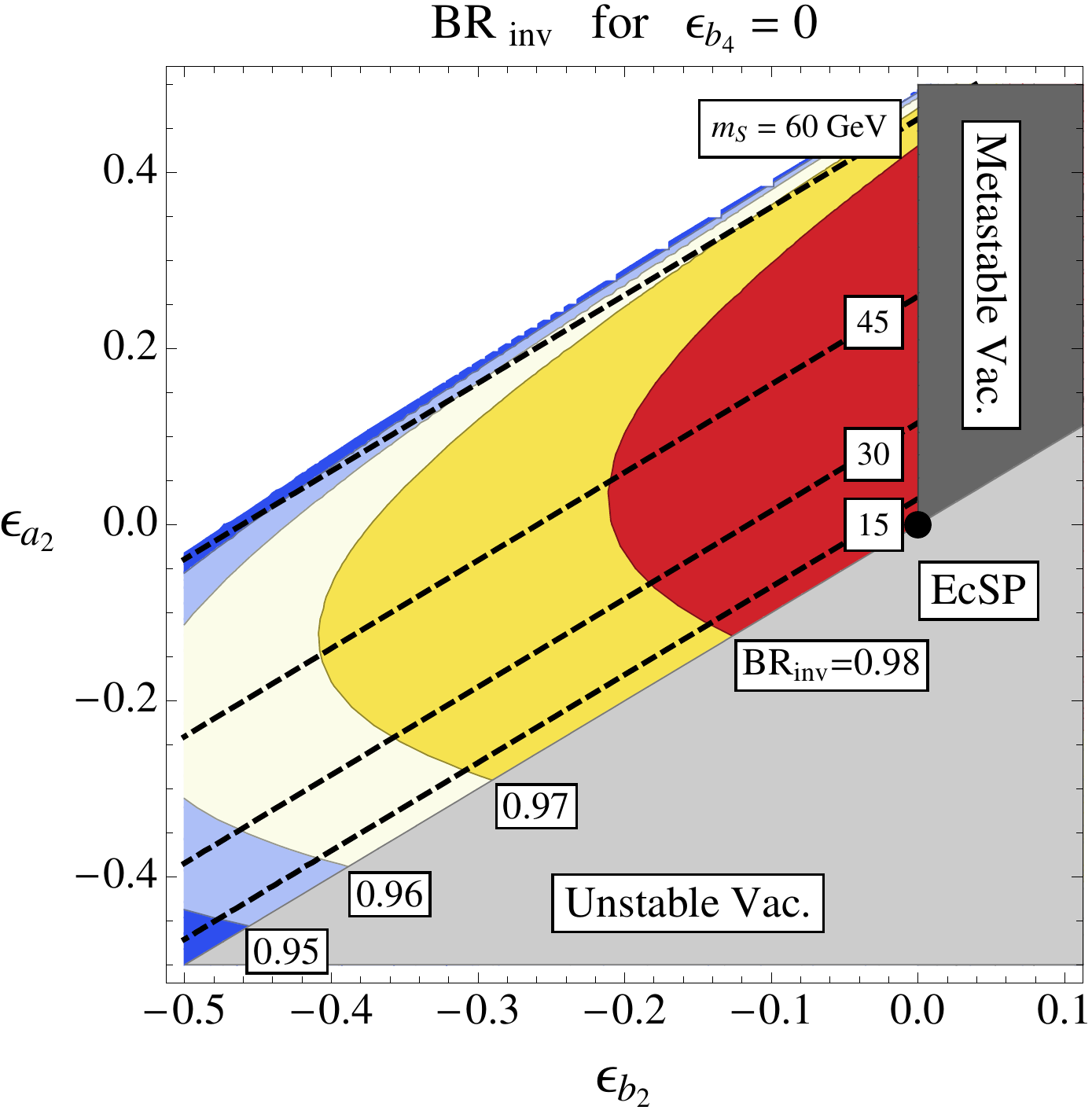} 
\caption{\label{fig:EcSP_neighborhood}
SFOPT correlated with large invisible decay in a neighborhood of the EcSP.  
The dashed lines corresponds to values of the singlet mass $m_S$.  
{\it Left.}  The EW order parameter $v(T_c) / T_c$, for which \eref{eq:vcoverTc_Z2xSM} is the leading-order expression.  
{\it Right.}  
The branching fraction of Higgs to an invisible S-pair ${\rm BR}_{\rm inv}$, for which the width \eref{eq:Z2xSM_Gammainv} is the leading-order expression (see also \cite{Barger:2008jx}).  
}
\end{center}
\end{figure}

We can proceed to perturb away from the EcSP by writing the parameters as
\begin{align}\label{eq:Z2xSM_deltaEcSP}
	b_2 = \mu^2 \left( 1 + \epsilon_{b_2} \right) 
	\ , \quad 
	b_4 = \lambda \left( 1 + \epsilon_{b_4} \right) 
	\ , \ \ {\rm and} \quad
	a_2 = 2 \lambda \left( 1 + \epsilon_{a_2} \right) \, .
\end{align}
What sort of perturbations will yield SFOPT?  
At the EcSP, the EW-broken and EW-symmetric vacua are degenerate, and if $c_h = c_s$ then the thermal corrections will maintain that degeneracy.  
As we perturb away from the EcSP looking for SFOPT, we will need to ensure that degeneracy breaking causes the EW-broken vacuum to be energetically favored and also ensure that the breaking of $c_h \neq c_s$ causes the EW-symmetric vacuum (in which $\mathbb{Z}_2$ is broken) to become (free-)energetically favored above some temperature.  
Keeping this picture in mind, we can proceed to calculate the phase transition parameters.  
In this neighborhood of the EcSP, the phase transition temperature and EW order parameter are given by 
\begin{align}
	T_c &= \frac{m_H}{2 \sqrt{c_h - c_s}} \sqrt{ \epsilon_{b_4} - 2 \epsilon_{b_2}} \Bigl( 1 +O(\epsilon_{b_2}, \epsilon_{b_4}) \Bigr)  \label{eq:Tc_Z2xSM} \\
	\frac{v(T_c)}{T_c} & = 2 \sqrt{c_h - c_s} \frac{v}{m_H}  \frac{1}{\sqrt{\epsilon_{b_4} - 2 \epsilon_{b_2}}} \Bigl( 1 + O(\epsilon_{b_2}, \epsilon_{b_4}) \Bigr) \, . \label{eq:vcoverTc_Z2xSM}
\end{align}
See also \fref{fig:EcSP_neighborhood}.  
As we anticipated, $T_c$ is arbitrarily small and $v(T_c) / T_c$ is arbitrarily large for arbitrarily small perturbations away from the EcSP ($\epsilon_{b_4} - 2 \epsilon_{b_2} \ll 1$).  
The particular combination of parameters $\epsilon_{b_4} - 2 \epsilon_{b_2}$ appears, because it controls the degree of degeneracy breaking between the EW-symmetric and EW-broken vacua.  
We can verify this by calculating
\begin{align}\label{eq:Delta_Veff}
	&V_{\rm eff}\left( 0, v_s , T \right)  - V_{\rm eff}\left( v , 0, T \right) \nn
	&\qquad = \Bigl[ \frac{\mu^4}{4 \lambda} \left(\epsilon_{b_4} - 2 \epsilon_{b_2} \right) - \frac{\mu^2}{2 \lambda} \left( c_h - c_s \right) T^2 \Bigr] \Bigl( 1+ O(\epsilon_{b_2} \sim \epsilon_{b_4} \sim T^2) \Bigr) \, ,
\end{align}
where $v_s = \sqrt{b_2 / b_4}$ is the expectation value of $s$ in the EW-symmetric vacuum and $v = \sqrt{\mu^2 / \lambda} = 246 \GeV$ is the Higgs {\sc vev}.  
Thus if $\epsilon_{b_4} - 2 \epsilon_{b_2} = 0$, the two vacua are degenerate at $T=0$.  
If $\epsilon_{b_4} - 2 \epsilon_{b_2} > 0$, the broken vacuum is energetically favored and the PT occurs at the temperature $T_c$ given by \eref{eq:Tc_Z2xSM}, but if $\epsilon_{b_4} - 2 \epsilon_{b_2} < 0$, the symmetric vacuum is energetically favored and the PT does not occur.  
From this discussion, and particularly \eref{eq:vcoverTc_Z2xSM}, we conclude that SFOPT are found in the neighborhood of the EcSP, but additionally the EcSP demarcates a boundary between physical models ($\epsilon_{b_4} - 2 \epsilon_{b_2} > 0$) in which EWSB occurs and unphysical models ($\epsilon_{b_4} - 2 \epsilon_{b_2} < 0$) in which EWSB does not take place.  
The singular factor of $1 / \sqrt{c_h - c_s}$ in \eref{eq:Tc_Z2xSM} can also be understood in light of \eref{eq:Delta_Veff}.  
If $c_h = c_s$, then thermal corrections lift the EW-broken and EW-symmetric phases together maintaining their degeneracy.  
One needs $c_h > c_s$ to ensure that $V_{\rm eff}$ at the EW-broken phase (free energy density) is lifted more greatly with increasing temperature than the EW-symmetric phase.  
Conversely, if $c_h < c_s$ then the EW-symmetric phase in which the $\mathbb{Z}_2$ is broken never becomes (free-)energetically favored.  

We can begin to investigate the phenomenology near the EcSP by calculating the mass of the singlet scalar field.  
The tree-level relationship can be read off of the Lagrangian \eref{eq:L_Z2xSM_EcSP}, which gives
\begin{align}
	m_S^2 = -b_2 + \frac{a_2}{2} v^2 \xrightarrow{\rm EcSP} \frac{m_H^2}{2} \left( \epsilon_{a_2} - \epsilon_{b_2} \right) \, .
\end{align}
Since this scalar field corresponds to the pseudo-Nambu-Goldstone boson of the spontaneously broken continuous symmetry, we are not surprised to find that it is light when deviations away from the EcSP are small $\epsilon_{a_2} - \epsilon_{b_2} \ll 1$.  
Note that at the EcSP, the Nambu-Goldstone boson kinetic term will couple to the Higgs through a dimension-five operator with a coupling strength of order $f/v$ where $f$ is a group theory factor typically of order unity.  
This generically leads to a large Higgs invisible\footnote{
More generally, $\mathbb{Z}_2$-violating couplings between the hidden sector and the SM may allow S to decay back into SM particles. ÊIn that case, the same constraints apply to the unobserved exotic decays.  
} decay width.  In the toy model at hand, the decay width is
\begin{align}\label{eq:Z2xSM_Gammainv}
	\Gamma(H \to SS) 
	& \xrightarrow{\rm EcSP} \frac{m_H^3}{32 \pi v^2} \Bigl( 1 + (\epsilon_{a_2} + \epsilon_{b_2}) + O(\epsilon_{a_2}^2, \epsilon_{b_2}^2) \Bigr) \, .
\end{align}
See also \fref{fig:EcSP_neighborhood}.  
Since $S$ only couples to the SM via the Higgs, the width for Higgs decay into SM fields, $\Gamma(H \to {\rm SM})$, is only affected by its coupling to $S$ at the multi-loop level.  
Thus, we can approximate $\Gamma(H \to {\rm SM})$ by the SM Higgs total width, which is $\Gamma_{\rm tot}^{\rm SM} \approx 5 \MeV$ for $m_H \approx 125 \GeV$ \cite{Djouadi:2005gi}.  
We find that the invisible branching ratio is
\begin{align}
	{\rm BR}_{\rm inv} = \frac{\Gamma(H \to SS)}{\Gamma(H \to {\rm SM}) + \Gamma(H \to SS)} \approx 0.985 \, ,
\end{align}
where we also neglect kinematically suppressed three-body (and greater) final states.  
Such a large invisible decay greatly exceeds the 95\% CL limits set by analyses of the LHC and Tevatron Higgs data, which were discussed in \sref{sec:DataandInterpretation}.  
Thus, the tension which we had discussed between the EcSP limit and invisible decay is illustrated in a concrete setting.  

We can attempt to evade the collider constraints on Higgs invisible decay by suppressing the channel $H \to SS$.  
This can be accomplished by moving away from the EcSP.
In the following, we will discuss two ways of 
deviating away from the EcSP while maintaining a SFOPT.  The first way
will be to reach an enhanced discrete symmetry point in the parameter
space such that the mass of the would-be Nambu-Goldstone boson is
lifted above the threshold for the two-body decay of the Higgs.  The
existence of a remnant symmetry is what guarantees the SFOPT in this
first deformation \cite{Barger:2011vm}.  The second way will be to
approach a free theory limit for the would-be Nambu-Goldstone boson
while maintaining a symmetry of the potential at the tree-level.  In
this second deformation, the kinetic term of the would-be
Nambu-Goldstone breaks the symmetry of the potential, but such
breaking is mild enough to ensure a SFOPT \cite{Barger:2011vm}.

Let us consider the first deformation.  Specifically, we perturb away
from the EcSP such that the continuous symmetry is broken to its
discrete subgroup $\mathbb{S}_2$ which exchanges $\sqrt{2} H
\leftrightarrow S$.  The EdSP is given by
\begin{align}\label{eq:EdSP}
	{\rm EdSP:} \qquad \left\{ b_2 = \mu^2 \quad , \quad b_4 = \lambda \right\} \quad \text{and} \quad \left\{g_i = 0 \quad , \quad y_i = 0 \right\} \, .
\end{align}
Since $\mu^2$, $\lambda$, and $a_2$ are free to vary, the EdSP
represents a three-dimensional submanifold of the full $\mathbb{Z}_2$xSM
parameter space in contrast to the two-dimensional submanifold
corresponding to EcSP.  As before we can consider perturbations away
from the EdSP parametrized as
\begin{align}\label{eq:Z2xSM_deltaEdSP}
	b_2 = \mu^2 \left( 1 + \epsilon_{b_2} \right) 
	\qquad {\rm and} \qquad
	b_4 = \lambda \left( 1 + \epsilon_{b_4} \right)  \, .
\end{align}
Because the singlet is no longer the Nambu-Goldstone boson of the
spontaneously broken continuous symmetry, its mass need not be small:
\begin{align}
	m_S^2 = -b_2 + \frac{a_2}{2} v^2 \xrightarrow{\rm EdSP} \frac{m_H^2}{4 \lambda} \left( a_2 - 2 \lambda \right) \left(1 - \frac{2 \lambda}{a_2 - 2\lambda}  \epsilon_{b_2} \right) \, .
\end{align}

From this expression we can see how the variation of $a_2$ affects the vacuum structure.  
For $a_2 = 2 \lambda$ we return to the EcSP and the singlet is the massless Nambu-Goldstone boson.  
For $a_2 < 2 \lambda$, the singlet becomes tachyonic, signaling that the true vacuum of the theory is one in which the $\mathbb{Z}_2$ is spontaneously broken.  
This is an undesirable limit, because without the $\mathbb{Z}_2$ preventing the Higgs and singlet from mixing, we run into the collider Higgs-mixing constraints, which were discussed in \sref{sec:DataandInterpretation}.  
For $a_2 > 2 \lambda$, the vacuum preserves the $\mathbb{Z}_2$ and the singlet is massive.  
Provided that $a_2 > 3 \lambda$, the singlet mass $m_S > m_H / 2$ will exceed the kinematic threshold and block the invisible decay $H \to SS$.  
Using $m_H \approx 125 \GeV$ and $\lambda = m_H^2 / (2 v^2)$, this bound is approximately $a_2 \gtrsim 0.39$.  
Moreover, since the expressions for the phase transition temperature and EW order parameter, \eref{eq:Tc_Z2xSM} and \eref{eq:vcoverTc_Z2xSM}, were independent of $a_2$, we still expect to find SFOPT in this corner of parameter space near the EdSP.  
Thus a departure from the EcSP along the EdSP allows for SFOPT while avoiding Higgs invisible decay by kinematically blocking the $H \to SS$ channel.  

A second deformation away from the EcSP while preserving SFOPT but avoiding Higgs invisible decay is obtained by moving towards a free theory ($a_2 = 0$) for the would-be Nambu-Goldstone field without making it heavy.  
To maintain the SFOPT, we must do this while preserving the degeneracy of the energy of the two vacua involved in the phase transition.  
In the previous discussion we saw that if we moved away from the EcSP along the direction of the EdSP, then taking $a_2 < 2 \lambda$ would lead to an undesirable change in the vacuum structure such that the $\mathbb{Z}_2$ becomes spontaneously broken and the Higgs and singlet are allowed to mix.  
Thus, we must find a different path that continuously connects the EcSP with $a_2 = 0$ but maintains the vacuum structure including the degeneracy.  

The correct path is given by the following parameter choice,
\begin{align}\label{eq:EcSPbar}
	\overline{\rm EcSP}: \qquad \left\{ b_2 = \frac{a_2}{2 \lambda} \mu^2 \quad , \quad b_4 = \left( \frac{a_2}{2 \lambda} \right)^2 \lambda \right\} \, .
\end{align}
At the parameter point \eref{eq:EcSPbar}, the scalar sector Lagrangian can be written as  
\begin{align}\label{eq:L_Z2xSM_EcSPbar}
	\mathcal{L}_{\mathbb{Z}_2 {\rm xSM}} \Bigr|_{\overline{\rm EcSP}} \supset &\left( \partial_{\mu} H \right)^{\dagger} \left( \partial^{\mu} H \right) + \frac{1}{2} \left( \partial_{\mu} S \right) \left( \partial^{\mu} S \right) \nn
	&- \left[ -\mu^2 \left( H^{\dagger} H + \frac{a_2}{2 \lambda} S^2 /2 \right) + \lambda \left( H^{\dagger} H + \frac{a_2}{2 \lambda} S^2/2  \right)^2 \right] \, .
\end{align}
From this expression we see that the scalar potential is invariant under a continuous symmetry transformation which rotates and dilates the fields $H$ and $S$, but that the scalar kinetic terms are not invariant (unless $a_2 = 2 \lambda$).  
Thus, \eref{eq:EcSPbar} is not a true enhanced symmetry point of the $\mathbb{Z}_2$xSM.  
Radiative corrections will spoil the symmetry, and therefore we do not expect the effective potential to respect this symmetry (even if we were to also set $g_i = y_i = 0$).  
Nevertheless, since in this class of models, the phase transition parameters are dominantly controlled by the structure of the tree-level scalar potential, we expect that SFOPT may still be found in the vicinity of \eref{eq:EcSPbar}.  
However, it turns out that in breaking this continuous symmetry the radiative corrections split the degeneracy of the EW-symmetric and EW-broken vacua in such a way that the EW-broken vacuum becomes metastable, and consequently EWSB does not occur.  
To avoid this outcome, we must allow for a finite perturbation away from the $\overline{\rm EcSP}$ parameter point.  
We consider instead the $\overline{\rm EcSP'}$ defined to be
\begin{align}\label{eq:Z2xSM_deltaEcSPbar}
	\overline{\rm EcSP'}: \left\{ \, b_2 = \frac{a_2}{2 \lambda} \mu^2 \left( 1 + \epsilon_{b_2} \right) \quad , \quad b_4 = \left( \frac{a_2}{2 \lambda} \right)^2 \lambda \left( 1 + \epsilon_{b_4} \right) \, \right\} \, ,
\end{align}
where we will allow $a_2$ to vary and keep $\epsilon_{b_2} =
\epsilon_{b_4} = -1/2$.

\begin{figure}[t]
\begin{center}
\includegraphics[width=0.65\textwidth]{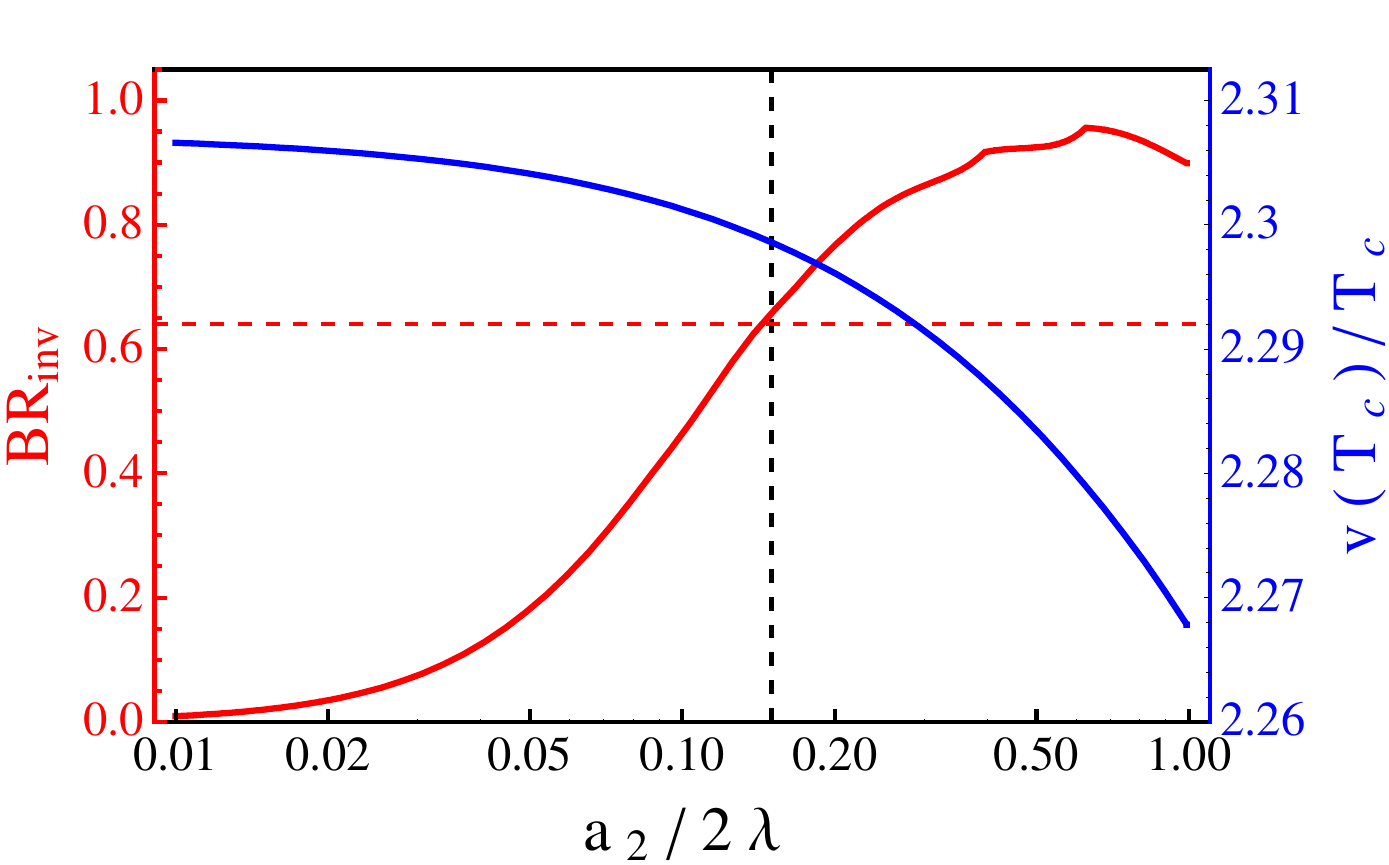}
\caption{\label{fig:Z2xSM_SFOPT}
The EW order parameter $v(T_c) / T_c$ (blue) and invisible branching fraction ${\rm BR}_{\rm inv}$ (red), calculated as in \fref{fig:EcSP_neighborhood} but at the $\overline{\rm EcSP'}$ parameter point \eref{eq:Z2xSM_deltaEcSPbar}.  
As $a_2 / 2 \lambda$ is decreased below about $0.15$, Higgs invisible decay becomes sufficiently suppressed to evade collider constraints which impose ${\rm BR}_{\rm inv} \lesssim 0.64$.  
This threshold corresponds to $a_2 \approx 0.043$.  
At the same time, the electroweak phase transition remains strongly first order $v(T_c) / T_c > 1$.  
}
\end{center}
\end{figure}

Along the trajectory \eref{eq:Z2xSM_deltaEcSPbar} we can take $a_2 \to
0$ while keeping $\lambda$ and $\mu$ finite.  The singlet remains
light $m_S^2 = - b_2 + a_2 v^2 / 2 = a_2 v^2 \epsilon_{b_2}$ and the
invisible width is approximately given by
\begin{align}\label{eq:GammaHSS_EcSPbar}
	\Gamma(H \to SS) 
	\xrightarrow{\overline{\rm EcSP'}} \frac{m_H^3}{32 \pi v^2} \left( \frac{a_2}{2 \lambda} \right)^2 \, .
\end{align}
To bring the invisible branching fraction below ${\rm BR}_{\rm inv} < 0.64$ (one of the weakest 95\% CL limits \cite{Espinosa:2012vu}) we need $a_2 < 0.043$.  
Furthermore, since the expression for the EW order parameter \eref{eq:vcoverTc_Z2xSM} is independent of $a_2$, we still expect to finds SFOPT in this limit.  
This can be verified by calculating the EW order parameter
numerically, and the result is shown in \fref{fig:Z2xSM_SFOPT}.  

In summary, within the Class IIA scenario, EcSP region of the SFOPT
parametric region is cleanly ruled out by the current data disfavoring
large Higgs branching to BSM states.  What is clear from the two
deformations away from EcSP in the context of a simple BSM model is
that EdSP does not require a small dimensionless parameter while
$\overline{\rm EcSP'}$ requires a tiny dimensionless coupling which
begs for an explanation.  In that sense, EdSP more naturally
accommodates both a SFOPT and the current Higgs data.

\subsection{Class IIB: Tree-Level (Non-Renormalizable Operators) Driven}\label{subsec:IIB}

The second way of obtaining a SFOPT using only tree-level operators is to employ nonrenormalizable terms in the potential.  
If the scale of new physics $\Lambda$ is not much larger than the EW scale, then the leading correction to the scalar potential, $(H^{\dagger} H)^3$, may dramatically change the nature of the EWPT.\footnote{
Here, we assume that the operator coefficient of the dimension-six Higgs kinetic term is vanishing.  More generally, a larger parameter space is consistent with SFOPT \cite{Grinstein:2008qi}.  }
In this scenario, the effective potential may be written as
\begin{align}\label{eq:CDrivenB_Veff}
	V_{\rm eff}(h, T) \approx \frac{1}{2} \left( \mu^2 + c \, T^2 \right) h^2 + \frac{\lambda}{4} h^4 + \frac{1}{8\Lambda^2} h^6 \per
\end{align}
Since typically $v(T_{c}) < v$, the $O(h^8 / \Lambda^4)$ terms can be neglected provided that $\Lambda > v$.   Ê
By minimizing the potential, the parameters $\mu^2$ and $\lambda$ may be exchanged for the Higgs {\sc vev} $v$ and mass $m_H$.  
These relationships are given by 
\begin{align}
	\lambda 
	&= \frac{m_H^2}{2 v^2} \left( 1 - \frac{\Lambda_{\rm max}^2}{\Lambda^2} \right)  \label{eq:Qdriven2_lambda} \\
	\mu^2 
	&= \frac{m_H^2}{2} \left(\frac{\Lambda_{\rm max}^2}{2 \Lambda^2} -1 \right) \label{eq:Qdriven2_musq} \com
\end{align}
where we have introduced $\Lambda_{\rm max} \equiv \sqrt{3} v^2 / m_H$, the meaning of which will become clear shortly.
Since we are interested in the limit that will yield a barrier in the effective potential, we will focus on the case of a low-scale cutoff such that $\mu^2 + c \, T^2 > 0$ stabilizes the EW-symmetric vacuum, $\lambda < 0$ causes the potential to turn over, and the $O(h^6)$ term stabilizes the EW-broken vacuum.  
In order to obtain $\lambda < 0$, we must have $\Lambda < \Lambda_{\rm max}$, where the upper bound evaluates to $\Lambda_{\rm max} \approx 800 \GeV$ for $m_H \approx 125 \GeV$.
Hence, if $h$ here is interpreted as exactly the Higgs direction such that $m_H$ is the Higgs mass, this class of models generically requires a low cutoff scale coming from trying to keep $v$ fixed and $\lambda<0$.  
As we will see, the consequent prediction of new states at the $800 \GeV$ scale is likely to be the strongest test of this class of scenarios.  
A potential of this form is illustrated in \fref{fig:Veff_plots}.  
The electroweak phase transition in this effective theory was studied in Refs.~\cite{Grojean:2004xa, Delaunay:2007wb, Grinstein:2008qi}.

As in the Class IIA scenario, the presence of the tree-level barrier allows $v(T_c) \approx v$ and therefore $v(T_c) / T_c$ may be enhanced by reducing $T_c$.  
Once again using standard techniques, we calculate the phase transition temperature and the EW order parameter to be
\begin{align}
	T_c & = \sqrt{ \frac{\mu^2}{c} } \sqrt {\frac{\lambda^2 \Lambda^2}{4 \mu^2} - 1 }	 \label{eq:ClassIII_Tc} \\
	\frac{v(T_c)}{T_c} & = \sqrt{ \frac{c}{- \lambda}} \frac{2}{\sqrt{1 - \frac{4 \mu^2}{\lambda^2 \label{eq:ClassIII_vTcoverTc}\Lambda^2} }} \, .
\end{align}
The optimal limits for enhancing $v(T_c) / T_c$ are given by the following.
\begin{description}
	\item[$\boldsymbol{c \gg \lambda}:$] \qquad This limit was
          discussed previously in the context of the Class IIA.
	\item[$\boldsymbol{4 \mu^2 / \lambda^2 \Lambda^2 \to 1}:$] \qquad
Using the relationships (\ref{eq:Qdriven2_lambda}) ands (\ref{eq:Qdriven2_musq}), this combination of parameters can be expressed as
\begin{align}\label{eq:lammax}
	\frac{4 \mu^2}{\lambda^2 \Lambda^2} = \frac{4}{3} \frac{1 - 2 \Lambda^2 / \Lambda_{\rm max}^2}{(1 - \Lambda^2/\Lambda_{\rm max}^2)^2} \per
\end{align}
Then, the limit is obtained when $\Lambda \to \Lambda_{\rm min}$ where $\Lambda_{\rm min} \equiv \Lambda_{\rm max} / \sqrt{3} = v^2 / m_H$.  
For $m_H \approx 125 \GeV$ this evaluates to $\Lambda_{\rm min} \approx 480 \GeV$.  
As we approach this limit, the phase transition temperature, given by \eref{eq:ClassIII_Tc}, goes to zero.  
We found a similar behavior in Class IIA, and once again we can identify this degeneracy limit with an EdSP \cite{Barger:2011vm}.

The $(H^{\dagger} H)^3$ operator is able to evade the standard phenomenological constraints.  
Since it preserves the custodial $\SU{2}$, there is no anomalous contribution to the $\rho$ parameter, even for a low cutoff \cite{Grojean:2004xa}.  
However, if other dimension-six operators are not forbidden, they may be constrained by electroweak precision tests.  
The Higgs cubic self-coupling, given by 
\begin{align}\label{eq:lamHHH}
	\lambda_{HHH} \equiv \frac{m_H^2}{v} \left( 1 + 2 \frac{\Lambda_{\rm min}^2}{\Lambda^2} \right) \com
\end{align}
receives $\ord{1}$ corrections in this limit.  
A measurement of $\lambda_{HHH}$ at the LHC is very difficult, but such large deviations from the SM have the potential to be measured with $1000 \, {\rm fb}^{-1}$ data at $14 \TeV$ \cite{Dolan:2012rv, Papaefstathiou:2012qe}.  

	\item[$\boldsymbol{\lambda \to 0}$] \qquad
We would like to take this limit $\lambda \to 0$ while fixing $4\mu^2 / (\lambda^2 \Lambda^2)$ such that \eref{eq:ClassIII_vTcoverTc} just scales like $1 / \sqrt{\lambda}$.  
Using the relationship \eref{eq:lammax}, this implies that we must let $\Lambda / \Lambda_{\rm max} = \const$~~ 
Then, \eref{eq:Qdriven2_lambda} reveals that in order to take $\lambda$ to zero we would have to take $m_H \propto \sqrt{\lambda}$ to zero.  
If $m_H$ is identified with the Higgs mass itself (recall that $h$ in principle can be a mixture of Higgs and another field direction), this limit is naively at tension with the Higgs mass not being much smaller than the electroweak scale, but -- as we will see below -- this is not necessarily a problem.
\end{description}

\begin{figure}[t]
\begin{center}
\includegraphics[width=0.45\textwidth]{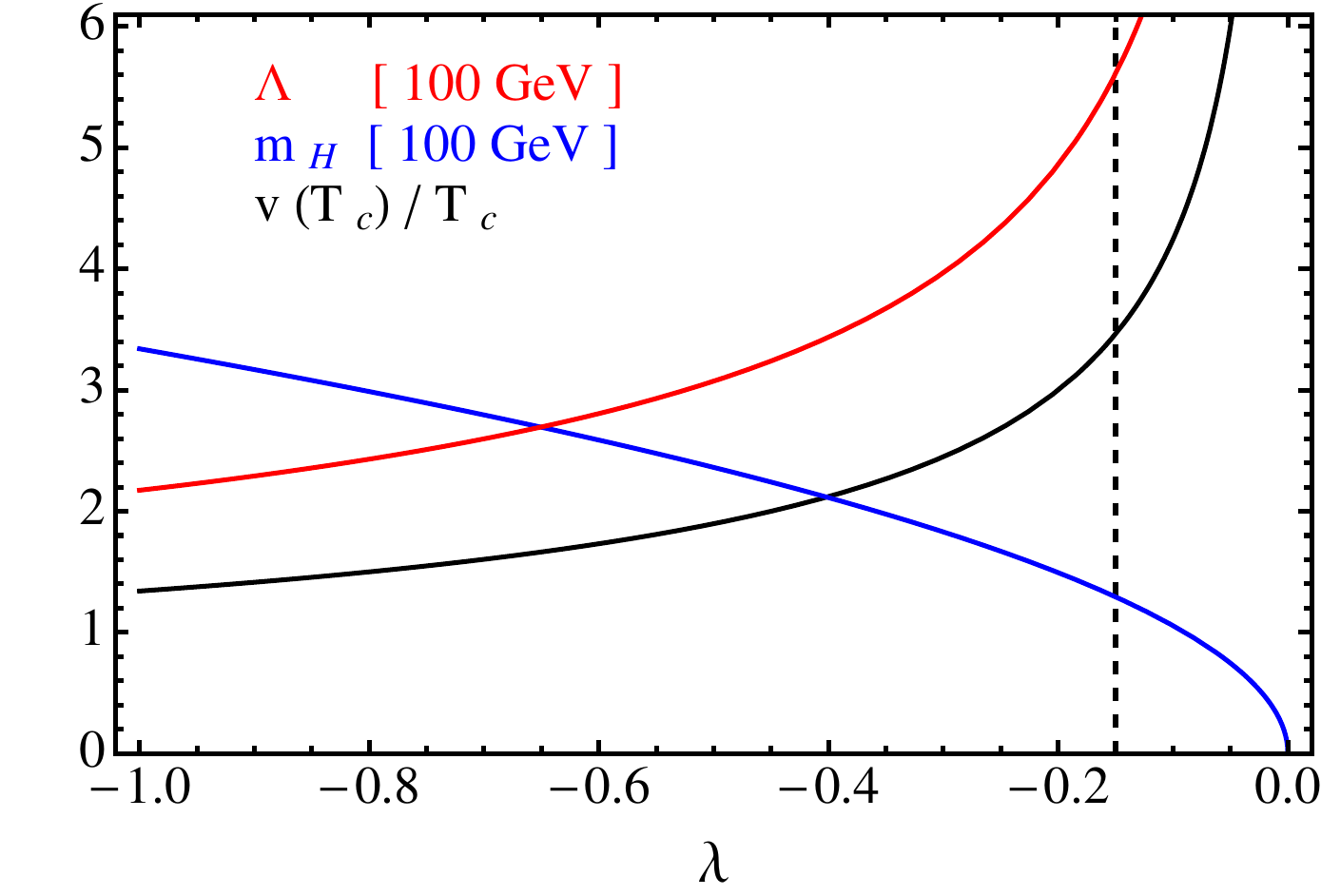} \hfill
\includegraphics[width=0.45\textwidth]{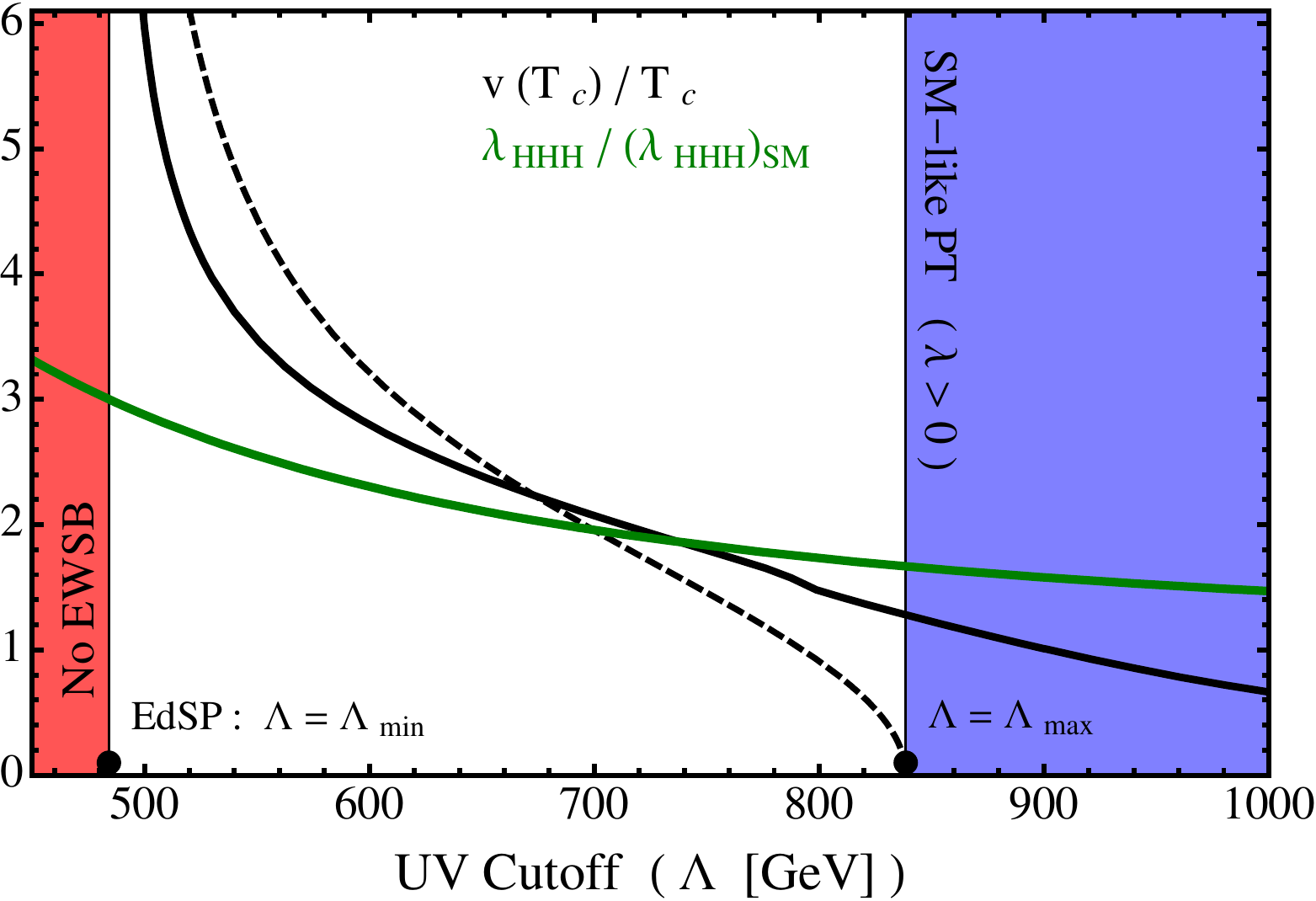} 
\caption{\label{fig:CDriven_NRops_EcSP}
{\it Left:}  The EWPT order parameter $v(T_c)/T_c$ \eref{eq:ClassIII_vTcoverTc} (black), Higgs mass $m_H$ (blue), and UV cutoff $\Lambda$ (red) as $\lambda$ is varied.  The parameters $m_H$ and $\Lambda$ are in units of $100 \GeV$.  
{\it Right:}  The EWPT order parameter with $m_H = 125 \GeV$ in the vicinity of the EdSP $\Lambda = \Lambda_{\rm min} \approx 480 \GeV$.  The solid black line shows the result of a calculation using the full one-loop thermal effective potential whereas the dashed line shows the approximation \eref{eq:ClassIII_vTcoverTc}.  The Higgs cubic self-coupling $\lambda_{HHH}$ (green) receives $O(1)$ corrections in the vicinity of the EdSP.  
}
\end{center}
\end{figure}

The nonrenormalizable $ (H^{\dagger}H)^3 $ term has been studied by \cite{Grojean:2004xa, Delaunay:2007wb} in the context of the electroweak phase transition and phenomenology.  
In their context, the $h$ of \eref{eq:CDrivenB_Veff} represents the SM Higgs field direction without mixing with another field degree of freedom.  
We have calculated $v(T_c) / T_c$ in the two limits (other than $c \gg \lambda$) discussed above.  
First, we allow $\lambda$ to vary while fixing 
\begin{align}\label{eq:nonEdSP}
	4 \mu^2 / \lambda^2 \Lambda^2 = 0.2 
\end{align}
(a value away from the EdSP).  
The results, shown in \fref{fig:CDriven_NRops_EcSP} (left panel), indicate that $v(T_c) / T_c$ grows like $1 / \sqrt{-\lambda}$ as $\lambda$ approaches zero.  
For $\lambda \approx -0.15$ the Higgs mass is consistent with the LHC signal at $m_H \approx 125 \GeV$, the phase transition is strongly first order, and the cutoff is low $\Lambda \approx 500 \GeV$, which may be problematic if the LHC does not discover new states at this energy scale.
On a positive note, as recognized by Refs.~\cite{Grojean:2004xa,Delaunay:2007wb}, this mechanism does not rely on there being small dimensionless couplings.  

The behavior of $v(T_c) / T_c$ near the EdSP is shown in \fref{fig:CDriven_NRops_EcSP} (right panel) where we have fixed $m_H = 125 \GeV$ and varied $\Lambda$.  
This figure illustrates that $v(T_c) / T_c$ grows as $\Lambda$ decreases toward the EdSP where $\Lambda_{\rm min} \approx 480 \GeV$ and $T_c$ vanishes.  
For smaller values of $\Lambda$, electroweak symmetry breaking does not occur (and the EWSB vacuum is metastable).  
For large values of the cutoff, the Higgs self-coupling $\lambda$ becomes positive and the PT proceeds as in the SM without any enhancement.  
Comparing the EdSP case to that of \eref{eq:nonEdSP}, we learn that because of the EdSP enhancement of the SFOPT, the maximum cutoff can be taken to be somewhat larger (although limited by $\Lambda_{\rm{max}}$), which may beneficial from a model-building perspective if data pushes the possibility of BSM states to higher energies.
The figure also shows that the Higgs cubic self-coupling $\lambda_{HHH}$ receives $O(1)$ corrections in the neighborhood of the EdSP; however, it will be difficult to measure this parameter at the LHC.

\subsection{Class III: Loop Driven}\label{subsec:III}
In the presence of qualitatively important quantum corrections, nonpolynomial field dependence may play a crucial role in rendering the electroweak phase transition strongly first order.  
For example, a competition between the terms $h^4$ and $h^4 \ln h^2$ may generate a barrier in the effective potential.
Alternatively, we can say that $\lambda$ is positive at high scales and runs negative at the electroweak scale.  
As a prototype of this ``Loop Driven'' model class, we will consider this running-quartic-coupling scenario whose effective potential may be written as
\begin{align}\label{eq:LDriven1_Veff}
	V_{\rm eff}(h, T) \approx \frac{1}{2} \left( \mu^2 + c \, T^2 \right) h^2 + \frac{\lambda}{4} h^4 + \frac{\kappa}{4} \, h^4 \ln \frac{h^2}{M^2} \, .
\end{align}
The parameters $\mu^2$ and $\lambda$ may be exchanged for
the Higgs {\sc vev} $v$ and mass\footnote{Since the loop contributions are
  important in this model class, we must be careful to distinguish the
  parameter $m_H$, defined as $m_H \equiv \sqrt{V_{\rm eff}^{\prime
    \prime}(v, T=0)}$, from the Higgs pole mass.  They differ by a
  correction that depends on the renormalization conditions.  Since we
  are primarily interested in the parametric scaling behavior and not
  numerical precision, we use $m_H$ to characterize the mass scale of
  fluctuations about $h = v$ and implement LHC Higgs data by setting
  $m_H \approx 125 \GeV$.  } $m_H$ using
\begin{align}
	\lambda & = \frac{m_H^2}{2 v^2} -\kappa \left( \ln \frac{v^2}{M^2} + \frac{3}{2} \right)  \\
	\mu^2 & = - \frac{m_H^2}{2} + \kappa \, v^2  \, .
\end{align}
The dimensionless parameter $\kappa$ parametrizes loop-suppressed corrections to the effective potential arising from interactions between the Higgs and the other fields.  
For example, in the SM one finds $\kappa_{\rm SM} \approx (6 M_W^4 + 3 M_Z^4 - 12 M_t^4) / (16 \pi^2 v^4) \approx  -0.018$.  
The loop-induced term can help provide a barrier -- as shown in \fref{fig:Veff_plots} -- if $\mu^2 > 0$ stabilizes $h = 0$, $\lambda < 0$ turns the potential over, and $\kappa > 0$ stabilizes $h = v$.  
To allow $\kappa > 0$, the BSM physics contributions should be dominated by bosonic fields, since fermion loops bring in an additional minus sign.  
Some examples of models that fall into this class are shown in Table \ref{tab:LDriven_Examples}.  

\begin{table}[t]
\hspace{-2in}
\vspace{-0in}
\begin{center}
\begin{tabular}{|c|p{7cm}|}
\hline
Model 
	& $-\Delta \mathcal{L}$
	\\ \hline \hline
Singlet \ Scalars \cite{Espinosa:2007qk, Espinosa:2008kw} 
	& $\sum_i^N  M^2 \abs{S_i}^2 + \lambda_S \abs{S_i}^4 + 2 \zeta^2 \abs{H}^2 \abs{S_i}^2$
	\\ \hline
Singlet \ Majoron \cite{Kondo:1991jz, Sei:1992np}
	& $\mu_s^2 \abs{S}^2 + \lambda_s \abs{S}^4 + \lambda_{hs} \abs{H}^2 \abs{S}^2 + \frac{1}{2} y_i S \nu_i \nu_i + \hc$
	\\ \hline
Two \ Higgs \ Doublets \cite{Cline:1996mga, Fromme:2006cm, Hambye:2007vf, Borah:2012pu} 
	& $\mu_D^2 D^{\dagger}D + \lambda_D (D^{\dagger}D)^2 + \lambda_3 H^{\dagger} H D^{\dagger}D + \lambda_4 \abs{H^{\dagger}D}^2 + (\lambda_5/2) [(H^{\dagger}D)^2 + \hc ]$ 
	\\ \hline
\end{tabular}
\end{center}
\caption{
\label{tab:LDriven_Examples}
Examples of models in the Loop Driven class. 
}
\end{table}%

The calculation of the EW order parameter from \eref{eq:LDriven1_Veff} requires the introduction of special functions (due to the nonpolynomial field dependence).  
A more transparent set of expressions is obtained by performing an expansion in $\epsilon = 1 - \kappa v^2 / m_H^2$, which we will see is a small quantity in the region of parameter space that turns out to be favorable for SFOPT.  
Doing so we obtain
\begin{align}
	T_c & \approx \frac{m_H}{2 \sqrt{c}} \sqrt{\epsilon} \left( 1 + \frac{1}{8} \epsilon + \frac{37}{384} \epsilon^2 + \ldots \right) \label{eq:QDriven_Tc} \\
	\frac{v(T_c)}{T_c} & \approx \frac{2 v \sqrt{c}}{m_H} \frac{1}{\sqrt{ \epsilon}} \left( 1 - \frac{3}{8} \epsilon - \frac{103}{384} \epsilon^2 + \ldots \right) \label{eq:QDriven_vTcoverTc}
\end{align}
The optimal limits for enhancing $v(T_c) / T_c$ are given by the following (recall $\epsilon = 1 - \kappa v^2 / m_H^2$).  
\begin{description}
	\item[$\boldsymbol{ \kappa v^2 / m_H^2 \to 1}:$]
In this limit, the quantum corrections are large, i.e., $\kappa \to \kappa_{\rm max} = m_H^2 / v^2 \approx 0.26$ for $m_H = 125 \GeV$.  
Since $\kappa$ contains a suppression factor of $1/16 \pi^2$, obtaining $\kappa = \ord{1}$ requires either many additional (bosonic) degrees of freedom and/or large couplings to the Higgs.  
This limit is then bounded by perturbativity constraints.  
Moreover, the large loops which generate $\kappa$ may also contribute to Higgs production and/or decay processes.  
We discuss this scenario further in \sref{sec:diphoton} in the context of Higgs diphoton decay.  
Finally, we can once again identify this limit as an EdSP in which $T_c$ vanishes and the EW-broken and EW-symmetric vacua are degenerate.  
Above $\kappa = \kappa_{\rm max}$ electroweak symmetry breaking does not occur.  
	\item[$\boldsymbol{ m_H \ll v \sqrt{c}}:$]
This limit is excluded in light of the Higgs discovery.  
\end{description}

As an example of a model in the Loop Driven class, we will discuss a singlet extension of the SM presented in \cite{Espinosa:2007qk}.  
The SM Lagrangian is extended by
\begin{align}
	\Delta \mathcal{L} = \sum_{i=1}^{N} \left( \partial S_i \right)^2 - \zeta^2 H^{\dagger} H \sum_{i=1}^{N} S_i^2 
\end{align}
where the $N$ real, scalar fields $S_i$ are singlets under the SM gauge group.  
We assume that $\zeta^2 > 0$ and the $S_i$ do not acquire {\sc vev}s.  
Instead, they modify the electroweak phase transition by radiatively generating a correction to the effective potential which is given by
\begin{align}
	\Delta V_{\rm eff} (h, T) = \frac{N \zeta^4 h^4}{64 \pi^2} \left[ \ln \frac{\zeta^2 h^2}{Q^2} - \frac{3}{2} \right] \, ,
\end{align}
when renormalized in the $\overline{\rm MS}$ scheme at the scale $Q$.  
This term can be matched onto the logarithmic term in \eref{eq:LDriven1_Veff} by choosing $\kappa = N \zeta^4 / 16 \pi^2$ and $M^2 = Q^2 \zeta^{-2} \ex{3/2}$.  
With this identification, the limit in which $\kappa v^2 / m_H^2$ approaches unity corresponds to $\zeta$ approaching $\zeta_{\rm max} = 2 \sqrt{\pi m_H / v} N^{-1/4}$, which evaluates to $\zeta_{\rm max} \approx 2.5 N^{-1/4}$ for $m_H \approx 125 \GeV$.  

\begin{figure}[t]
\begin{center}
\includegraphics[width=0.60\textwidth]{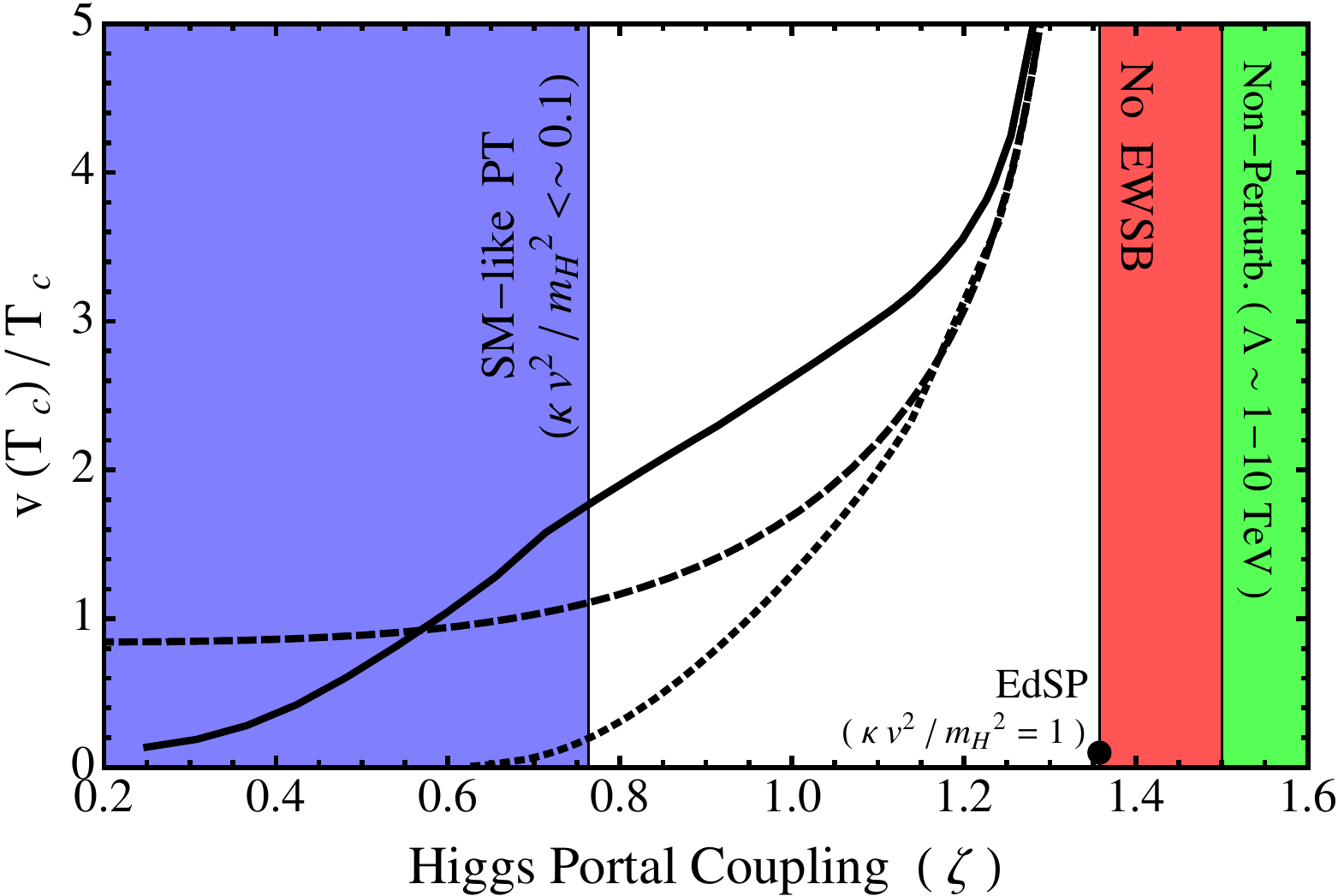} 
\caption{\label{fig:QDriven_Espinosa}
The EW order parameter evaluated (i) using the approximation \eref{eq:QDriven_vTcoverTc} (dashed), (ii) using the toy model potential \eref{eq:LDriven1_Veff} but without any further approximation (dotted), and (iii) using the full one-loop thermal effective potential, as described in the text (solid).  All three calculations reveal that $v(T_c) / T_c$ grows upon approaching the EdSP at $\zeta_{\rm max} \approx 1.36$.  For a sufficiently low cutoff ($\Lambda \sim 1 - 10 \TeV$), perturbativity is maintained up to $\zeta \sim 1.5$ \cite{Espinosa:2007qk}.  
}
\end{center}
\end{figure}

Choosing $m_H = 125 \GeV$, $N = 12$, and $Q = m_t = 172 \GeV$, we calculate $v(T_c) / T_c$ using the full one-loop thermal effective potential, and we present the results in \fref{fig:QDriven_Espinosa}.  
As expected, $v(T_c) / T_c$ grows upon approaching the EdSP where $\kappa v^2 / m_H^2 = 1$ corresponds to $\zeta_{\rm max} \approx 1.36$.  
For larger values of $\zeta$, electroweak symmetry breaking does not occur.  
For sufficiently small values of $\zeta$, the PT becomes SM-like and no longer strongly first order.  
The discrepancy between the approximations and the full one-loop calculation of $v(T_c) / T_c$ can be attributed to the implicit use of the high-temperature expansion in \eref{eq:LDriven1_Veff} and setting $c = c_{\rm SM} \approx 0.36$ without accounting for the $S$ field.  
The contribution from $S$ is suppressed at large $\zeta$ (where the approximation agrees well), because $S$ is heavy and its thermal contribution is Boltzmann-suppressed.  
At smaller values of $\zeta$, the $S$ field effectively renders $c > c_{\rm SM}$, which tends to increase $v(T_c) / T_c$, as indicated by \eref{eq:QDriven_vTcoverTc} and confirmed by \fref{fig:QDriven_Espinosa}.  
In this model, the additional singlet scalars will not have an appreciable impact on collider physics.  
We discuss the more general case of charged or colored scalars in \sref{sec:diphoton}.  

\section{Diphoton Excess and SFOPT in the Higgs Portal}\label{sec:diphoton}

One tantalizing hint of new physics in the recent LHC announcement is the observed excess of events in the final states with two photons.  
The $\gamma \gamma$ final state, which is associated with Higgs production by gluon fusion, is observed at a rate that exceeds the SM prediction by a factor of approximately $1.5$, while the $\gamma \gamma jj$ final state is enhanced by a factor of approximately three \cite{Giardino:2012dp}.  
Although not statistically significant yet, fits to the entire data set seem to favor an enhancement of the diphoton decay rate $\Gamma(h \to \gamma \gamma)$ by a factor of approximately $2-3$ with respect to the SM prediction, as well as a suppression of the gluon fusion production cross section $\sigma(gg \to h)$ by a factor of approximately $0.5-0.6$ (see, e.g., \cite{Buckley:2012em, Giardino:2012dp} and references therein).

Since $gg \to h$ and $h \to \gamma \gamma$ are both loop-induced processes in the SM, these channels are particularly sensitive to new physics.  
For instance, the appropriate enhancement and suppression can be achieved by letting the Higgs couple to a new scalar $S$ via the Higgs portal \cite{Batell:2011pz, Chang:2012ta, An:2012vp}.  
If $S$ is charged, graphs containing an $S$ loop will contribute to the amplitude for $h \to \gamma \gamma$ and interfere with the $t$ and $W$ loops that dominant the SM contribution.  
Generally, {\it a negative value of the Higgs portal coupling is favored} if the $h \to \gamma \gamma$ rate is enhanced, because then the $S$ loop will interfere constructively with the SM contribution.  
Furthermore, if $S$ is colored it will also interfere destructively with the SM $gg \to h$.  
As we have seen, the Higgs portal operator also provides a means of rendering the electroweak phase transition strongly first order.  
It is then interesting to ask whether the region of parameter space that can accommodate a SFOPT can also allow for enhanced diphoton decay in such simple models where a single Higgs portal
operator is responsible for both phenomena.  
We will see that generically, {\it the SFOPT condition favors a positive value of the Higgs portal coupling} and, therefore, is at tension with the diphoton enhancement in such minimal model settings.

In order to demonstrate that SFOPT favors a positive Higgs portal coupling, let us consider such an interaction between the Higgs and a scalar field $S$, as given by the Lagrangian
\begin{align}\label{eq:HiggsPortal}
	- \mathcal{L} \supset \mu_S^2 S^{\ast} S + 2 \lambda H^{\dagger} H S^{\ast} S \per
\end{align}
The phase transition calculation is independent of the quantum numbers of $S$ at the one-loop order, but instead only depends upon the coupling of $S$ to the Higgs.\footnote{If $S$ is colored, then the two-loop contribution from gluons can have an appreciable impact on the order of the phase transition.  This is, for example, the case in the MSSM \cite{Carena:1996wj}.  }
However, in order to obtain an enhanced diphoton decay rate, we need $S$ to carry an electric charge.  
Consequently, we must ensure that $S$ does not acquire a {\sc vev}.\footnote{This discussion presumes that $S$ is a singlet under weak isospin.  More generally, the electrically neutral component of $S$ may acquire a {\sc vev} without breaking $\U{1}_{\rm em}$.  However, unless this {\sc vev} is much less than $v$, it will be at tension with electroweak precision measurements.  }
In that case, the field-dependent squared mass of the $S$ field is given by 
\begin{align}
	m_{{\rm eff}, S}^2(h,T) = \mu_S^2 + \lambda h^2 + \Pi_S(T) \com
\end{align}
where $\Pi_S(T)$ is the thermal self-energy correction.  
In the appropriate limits, this simple extension of the SM Lagrangian can yield any one of the phase transition model classes discussed above.  
These are as follows.  
\begin{description}
	\item[\qquad {\bf Class I.  Thermally (BEC) Driven.}]  { The BEC term receives a contribution $(\mu_S^2 + \lambda h^2 + \Pi_S(T))^{3/2}$.  As discussed in \sref{subsec:I}, we must tune $\mu_S^2 \approx - \Pi_S(T_c)$.  However, in this limit the mass of the $S$ field is $m_{\rm eft, S}^2(v,0) = -\Pi_S(T_c) + \lambda v^2$.  We cannot let $\lambda < 0$, because this would render $S$ tachyonic and induce a {\sc vev}.  }
	\item[\qquad {\bf Class IIA.  Tree-Level (Ren. Ops.) Driven.}]  Since $S$ cannot acquire a {\sc vev}, the only way in which tree-level terms can enhance the strength of the phase transition are if $S$ had a nonzero expectation value in the early universe which returned to zero during the electroweak phase transition.  This scenario is realized by letting $\mu_S^2 < 0$ such that $S$ obtains a nonzero expectation value in the early universe, but ensuring that $-\lambda v^2 <\mu_S^2 < 0$ such that $S$ has a vanishing {\sc vev} today.  Once again, we find that $\lambda > 0$ is required for a SFOPT in this model class as well.  
	\item[\qquad {\bf Class IIB.  Tree-Level (Non-Ren. Ops.) Driven.}]  The nonrenormalizable operator $(H^{\dagger} H)^3 / \Lambda^2$ may be generated by integrating out the field $S$.  
The leading-order contribution to this operator coefficient is proportional to $(1/16 \pi^2) (\lambda^3 / M_S^2)$.  
Since this model class relies upon $(H^{\dagger} H)^3$ having a positive coefficient in order to stabilize the potential against a runaway direction, we must take $\lambda > 0$.  
	\item[\qquad {\bf Class III.  Loop Driven.}]  This model class relies upon the addition of a term to the effective potential that goes like $h^4 \ln h^2$ and its competition with the $h^4$ term to generate a barrier in the effective potential.  The Higgs portal operators \eref{eq:HiggsPortal} will instead generate a term of the form $h^4 \ln (\mu_S^2 + \lambda h^2)$.  Unless $\abs{\mu_S^2} \ll \abs{\lambda v^2}$, this term will simply scale like $h^4$ and there will be no competition between terms and no barrier.  However, if $\lambda < 0$, then in this limit the $S$ field develops a tachyonic instability and acquires a {\sc vev}.  
\end{description}

This analysis may seem to suggest that $\lambda > 0$ is generally favored by SFOPT.  
However, this is not the case.  
If we were not interested in enhancing Higgs diphoton decay, then we could achieve a SFOPT by coupling the Higgs to a singlet scalar field using the operators \eref{eq:HiggsPortal}.  
Choosing $\lambda < 0$, there exist models in the Tree-Level (Ren. Ops.) Driven class which achieve an SFOPT when the singlet has a {\sc vev}, which is restricted by Higgs-mixing constraints (see, e.g., \cite{Profumo:2007wc}).  
Furthermore, in a nonminimal-model setting in which we introduce additional singlet and charged scalars, the singlet(s) can enhance the phase transition while the charged field(s) can enhance the diphoton rates.

\section{Conclusion}

In this paper we have proposed a classification of the electroweak symmetry breaking sector which may yield a strongly first order phase transition -- a necessary ingredient for electroweak baryogenesis.
For each model class, we assumed that the last phase transition associated with the electroweak symmetry breaking sector was an electroweak symmetry breaking transition (i.e.,~no broken vacuum to broken vacuum transition), and we investigated the impact of the data that is currently available from the LHC: (i) the discovery of a $125 \GeV$ Higgs-like scalar, (ii) the absence of a large exotic (e.g., invisible) decay width, and (iii) the absence of a universal suppression, which would indicate mixing between the Higgs and a hidden sector scalar field.  
We find that the mass measurement severely constrains models (such as the MSSM \cite{Curtin:2012aa, Cohen:2012zz}) which drive a strongly first order phase transition with thermal loop effects.  
The invisible decay and mixing constraints are at tension with models which rely on light singlets coupled to the Higgs.

One recurring theme of our analysis is the ubiquity of enhanced symmetry points.  
We find that the ``optimal'' limit for SFOPT often corresponds to a parameter point at which the symmetry group of the theory is extended.  
In the case that the group is enlarged by a continuous symmetry, either the Higgs mass constraint or the  exotic decay and mixing constraints will come into play.  
The case of a discrete symmetry is less restricted \cite{Barger:2011vm}.

We have also discussed the possibility of employing the same Higgs portal operator to both render the EWPT strongly first order and to account for the diphoton excess observed by ATLAS and CMS.  
We find that these two goals are at odds with one another in the minimal model setting: the phase transition favors a positive Higgs portal coupling whereas the diphoton enhancement favors a negative coupling.  
A model which can accommodate EWBG as well as fit the LHC data will most likely require two distinct new-physics operators.  
However, it is worth noting that the diphoton excess does not have a great statistical significance, and the data remains consistent with the SM at the 75 \% CL \cite{Carmi:2012in} or approximately $2 \sigma$ \cite{Espinosa:2012im, Giardino:2012dp}.  
It is still entirely possible that the particle recently discovered by ATLAS and CMS is the SM Higgs \cite{Ellis:2012hz}.

\begin{acknowledgments}
DJHC and AJL were supported in part by the DOE through grant DE-FG02-95ER40896.  
LTW is supported by the NSF under grant PHY-0756966 and the DOE Early Career Award under grant DE-SC0003930.  
\end{acknowledgments}

\newpage
\bibliographystyle{JHEP}
\bibliography{refs_SFOPTclasses}

\end{document}